\documentstyle[epsfig,12pt]{article}
\newcommand{\be}{\begin{eqnarray}}
\newcommand{\ee}{\end{eqnarray}}
\newcommand{\ba}{\begin{array}}
\newcommand{\ea}{\end{array}}
\newcommand{\ur}[1]{(\ref{#1})}
\newcommand{\urs}[2]{(\ref{#1},\ref{#2})}
\renewcommand{\vec}[1]{{\bf #1}}
\newcommand{\eq}[1]{eq.~(\ref{#1})}
\newcommand{\eqs}[2]{eqs.(\ref{#1}, \ref{#2})}
\newcommand{\Eq}[1]{Eq.~(\ref{#1})}

\newcommand{\IIs}{$I$'s and $\bar I$'s }

\newcommand{\tr}{{\rm tr}\,}
\newcommand{\Tr}{{\rm Tr}\,}

 \def\Sp{\mbox{Sp}}

 \def\Dirac#1{#1\hskip-6pt/}
 \def\dd{\Dirac\partial}
 \def\dpp{\Dirac p}
 \def\dk{\Dirac k}

  \newcommand{\la}[1]{\label{#1}}
  
  \def\beq{\begin{equation}}
  \def\eeq{\end{equation}}
  \def\beqr{\begin{eqnarray}}
  \def\eeqr{\end{eqnarray}}
  
 \def\Dirac#1{#1\hskip-6pt/}
 \def\dd{\Dirac\partial}

\def\Ug{{U^{\gamma_5}}}

\begin{document}
\vspace{.8cm}
\begin{center}
{\Large\bf Chiral Quark-Soliton Model \footnote{Lectures
at the Advanced Summer School on Non-Perturbative Field Theory,
Peniscola, Spain, June 2-6, 1997.}} \\[1cm]

{\large\bf Dmitri Diakonov}
\\[.5cm]
{\it NORDITA, Blegdamsvej 17, 2100 Copenhagen \O, Denmark}\\
and\\
{\it Petersburg Nuclear Physics Institute, Gatchina,
St.Petersburg 188350, Russia \\}
\end{center}
\vskip .8true cm
\begin{abstract}
\noindent
The Chiral Quark-Soliton Model of nucleons is based on two ideas:
1) the major role of spontaneous chiral symmetry breaking in
hadron physics and 2) the relevance of the large $N_c$ (= number
of colours) limit for the real world. In these lectures I review
the theoretical foundations of the model, the physics involved, and
some of applications.
\end{abstract}
\vskip .8true cm
\tableofcontents
\vspace{.5cm}

\section{How do we know chiral symmetry is spontaneously broken?}
\setcounter{equation}{0}

The QCD lagrangian with $N_f$ massless flavours is known to
posses a large global symmetry, namely a symmetry under $U(N_f)\times
U(N_f)$ independent rotations of left- and right-handed quark fields.
This symmetry is called {\em chiral} \footnote{The word was coined by
Lord Kelvin in 1894 to describe moleculas not superimposable on its
mirror image.}. Instead of rotating separately the 2-component Weyl
spinors corresponding to left- and right-handed components of
quark fields, one can make independent vector and axial $U(N_f)$
rotations of the full 4-component Dirac spinors -- the QCD lagrangian
is invariant under these transformations too.

Meanwhile, axial transformations mix states with different
P-parities.  Therefore, were that symmetry exact, one would observe
parity degeneracy of all states with otherwise the same quantum
numbers.  In reality the splittings between states with the same
quantum numbers but opposite parities are huge. For example, the
splitting between the vector $\rho$ and the axial $a_1$ meson is $(1260
- 770)\simeq 500\;MeV$; the splitting between the nucleon and its
parity partner is even larger:  $(1535 - 940)\simeq 600\; MeV$.

The splittings are too large to be explained by the small bare or
current quark masses which break the chiral symmetry from the
beginning. Indeed, the current masses of light quarks are: $m_u \simeq
4\;MeV,\;\;m_d\simeq 7\;MeV,\;\;m_s\simeq 150\;MeV$. The only
conclusion one can draw from these numbers is that the chiral symmetry
of the QCD lagrangian is broken down {\em spontaneously}, and very
strongly. Consequently, one should have light (pseudo) Goldstone
pseudoscalar hadrons -- their role is played by pions which indeed are
by far the lightest hadrons.

The order parameter associated with chiral symmetry breaking is
the so-called {\em chiral} or {\em quark condensate}:

\beq
\langle\bar\psi\psi\rangle\simeq -(250\;MeV)^3.
\la{chcond}\eeq
It should be noted that this quantity is well defined only for
massless quarks, otherwise it is somewhat ambigious. By definition,
this is the quark Green function taken at one point; in momentum space
it is a closed quark loop. If the quark propagator has only the `slash'
term, the trace over the spinor indices implied in this loop would give
an identical zero. Therefore, chiral symmetry breaking implies that
a massless (or nearly massless) quark develops a non-zero dynamical
mass (i.e. a `non-slash' term in the propagator). There are no reasons
for this quantity to be a constant independent of the momentum;
moreover, we understand that it should anyhow vanish at large momentum.
The value of the dynamical mass at small virtuality can be estimated as
one half of the $\rho$ meson mass or one third of the nucleon mass,
that is about

\beq
M(0)\simeq 350-400\;MeV;
\la{M0}\eeq
this scale is also related to chiral symmetry breaking and should
emerge together with the condensate \ur{chcond}.

One could imagine a world without confinement but with chiral symmetry
breaking: it would not be drastically different from what we meet in
reality. There would be a tightly bound light Goldstone pion, and
relatively loosely bound $\rho$ meson and nucleon with
approximately correct masses, which, however, would be possible to
`ionize' from time to time.  Probably the spectrum of the highly
excited hadrons would be wrong, though even that is not so clear
\cite{DP6}.  We see, thus, that the spontaneous chiral symmetry
breaking is the main dynamical happening in QCD, which determines the
face of the strong interactions world.

If one understands the microscopic mechanism of spontaneous chiral
symmetry breaking and knows how to get the quantities \urs{chcond}{M0}
from the only dimensional parameter there is in massless QCD, namely
$\Lambda_{{\rm QCD}}$, one gets to the heart of hadron physics. My sense
is that it is achieved by ways of the QCD instanton vacuum, see
\cite{D1,SS} for recent reviews. Therefore, I start by showing in
section 2 how instantons lead to a low-energy theory exhibiting chiral
symmetry breaking and the appearance of a momentum-dependent
constituent quark mass $M(k)$. In section 3 the resulting effective
chiral lagrangian is theoretically studied and in section 4 it is
applied to build the Chiral Quark-Soliton Model.

\section{Low-energy limit of QCD from instantons}
\setcounter{equation}{0}

The idea that the QCD partition function is dominated
by instanton fluctuations of the gluon field, with quantum
oscillations about  them, has successfully confronted the majority of
facts we know about the hadronic world (for a review see
ref.\cite{SS}).  Instantons have been reliably identified in lattice
simulations (for a review see ref.\cite{vB}), and their relevance to
hadronic observables clearly demonstrated \cite{Negele-L97}
\footnote{It can be added that in the solvable $N=2$
supersymmetric version of QCD it is instantons --- and nothing besides
them --- that seem to be sufficient to reproduce the expansion of the
exact Seiberg--Witten prepotential \cite{KhozeMattisDorey}.}.
Thus, the effective low-energy theory coming from instantons
seems to be well-motivated. The aim of this section is to derive
this effective theory to which QCD is reduced at low momenta.

\subsection{Some of the results}

There is a well-known general statement that to get chiral symmetry
breaking one needs a finite spectral density $\nu(\lambda)$ of the
quark Dirac operator at zero eigenvalues, since the chiral
condensate is proportional to exactly this quantity:
$\langle\bar\psi\psi\rangle=-\pi\nu(0)/V^{(4)}$ \cite{BC}. A natural
way to get $\nu(0)\ne 0$ is to have a finite density of instantons and
antiinstantons (\IIs for short) in the 4-dimensional space-time,
$N/V^{(4)}$. Indeed in the presence of the topologically non-trivial
gluon fluctuations fermions necessarily have an exact zero mode
\cite{tH}, as it follows from the Atiah--Singer index theorem.
In the ensemble of \IIs the would-be zero modes in the background field
of individual \IIs are smeared into a band with a finite spectral
density at zero eigenvalues \cite{DP2}, leading to
$\nu(0)\ne 0$. The instanton vacuum provides thus a beautiful mechanism
of chiral symmetry breaking \cite{DP3}.

There are two mathematically equivalent ways to treat quarks in
the instanton vacuum. One is to calculate an observable in a given
instanton backgound and then to average over the collective coordinates
of \IIs and sum over their total numbers, $N_+$ and $N_-$. This
approach has been developed in refs. \cite{DP3,Pob}. The quark
propagator in the instanton vacuum takes the form of a massive
propagator with a dynamically generated momentum-dependent mass (in
Euclidean space, hence the factor $i$ in the `non-slash' term):

\beq
S(p)=\frac{\dpp+iM(p^2)}{p^2+M^2(p^2)},\;\;\;\;
M(p^2)={\rm const}\cdot\sqrt{\frac{N\pi^2\bar\rho^2}{VN_c}}
F^2(p\bar\rho),
\la{qprop}\eeq
Here $N/V$ is the instanton density at equilibrium and
$\bar\rho$ is the average instanton size, $F(p\bar\rho)$
is a combination of modified Bessel functions and is related to the
Fourier transform of the would-be zero fermion mode of individual
instantons,

\beq
F(p\rho)=2z\left[I_0(z)K_1(z)-I_1(z)K_0(z)
-\frac{1}{z}I_1(z)K_1(z)\right]_{z=p\rho/2}
\mathrel{\mathop{\longrightarrow}\limits_{p\rightarrow\infty}}
\frac{6}{(p\rho)^3},\;\;\;\;\;F(0)=1.
\la{Ff}\eeq
The numerical constant in \eq{qprop} is of the order of unity and
is determined by the self-consistency or gap equation:

\beq
4N_c\int \frac{d^4 p}{(2\pi)^4}\frac{M^2(p)}{M^2(p)+p^2}
=\frac{N}{V}.
\la{selfcons}\eeq

The chiral condensate $\langle\bar\psi\psi\rangle$ is the
quark propagator taken at one point; in momentum space it is
given by a quark loop:

\beq
-\langle\bar\psi\psi\rangle_{Mink}=i\langle\psi^\dagger\psi\rangle_{Eucl}
\approx 4N_c\int \frac{d^4 p}{(2\pi)^4}\frac{M(p)}{M^2(p)+p^2}
={\rm const^\prime}\cdot\sqrt{\frac{NN_c}{V\pi^2\bar\rho^2}}.
\la{chircond}\eeq

To get the numerical estimates of the condensate and of the
constituent quark mass one may rely on the variational
calculation of the instanton vacuum characteristics \cite{DP1,DPW},
which relates them to the only dimensional parameter in QCD,
$\Lambda_{QCD}$. Taking $\Lambda_{\overline{MS}}^{(3)} = 280\,MeV$,
one finds from refs. \cite{DP1,DPW} the basic characteristics
of the instanton vacuum, namely the average distance between
neighbouring instantons, $\bar R \equiv (N/V)^{-1/4}$ and their
average mean square radius, $\bar\rho$, to be

\beq
\bar R\approx 1\,fm,\;\;\;\;\; \bar\rho \approx 0.35\,fm.
\la{densrho}\eeq
Using these basic quantities one gets from \eqs{selfcons}{chircond}:

\beq
M(0)\approx 350\,MeV,\;\;\;\;\; -\langle\bar\psi\psi\rangle
\approx (250\,MeV)^3.
\la{Mchcon}\eeq
Another quantity closely associated with chiral symmetry breaking
is the pion decay constant which in the instanton vacuum is given
by \cite{DP3}

\beq
F^2_\pi\approx
4N_c\int \frac{d^4 p}{(2\pi)^4}\frac{M^2(p)}{[M^2(p)+p^2]^2}
={\rm const^{\prime\prime}}\cdot\frac{N}{V}\,
\bar\rho^2\,\ln\left(\frac{\bar R}{\bar\rho}\right) \simeq
(100\,MeV)^2.
\la{Fpi}\eeq

All these quantities appear to be close to their phenomenological
values. I would say that I don't know of any other approach to
non-perturbative QCD (except, of course, brute-force lattice
calculations) which would relate observables directly to
$\Lambda_{QCD}$, and with such an accuracy. Personally, I conclude that
the idea that the (Euclidean) QCD partition function is saturated
by relatively dilute instantons with quantum fluctuations of
gluon field about them, works quite satisfactory.

\subsection{Instanton-induced interactions}

As I have mentioned, these results have been obtained from
considering the motion of light quarks in a given instanton
background and then averaging it over the instanton ensemble
\cite{DP3}. There is a mathematically equivalent technique to
rederive these results, namely one first averages over the
instanton ensemble \cite{DP4}. This averaging induces many-quark
interactions whose simplified version was first suggested by
't Hooft \cite{tH}. Using the effective quark interaction theory
one can calculate various observables. According to the derivation
of ref. \cite{DP4} (recently reviewed in refs. \cite{D1,DPW})
averaging over instanton ensemble leads to a specific form
of the QCD partition function valid at low momenta, $p\le 1/\bar\rho$.
In what follows we shall use the Euclidean formulation of the theory;
$N_f$ is the number of light fermion flavours whose masses are put to
zero for simplicity. The effects of non-zero current quark masses
have been considered in ref. \cite{DPW}.

It is convenient to decompose the 4-component Dirac bi-spinors
describing quark fields into left- and right-handed Weyl spinors
which we denote as

\beq
\psi^{f\alpha i}_{L(R)},\;\;\;\;\;\psi^\dagger_{L(R)f\alpha i},
\la{psi}\eeq
where $f=1...N_f$ are flavour, $\alpha=1...N_c$ are colour
and $i=1,2$ are spinor indices. Let us introduce the
't Hooft-like $2N_f$-fermion vertices generated by \IIs,
which we denote by $Y_{N_f}^{(\pm)}$, respectively. These vertices
are obtained by explicit averaging over (anti)instanton orientation
matrices $U^\alpha_i$ and over the instanton size distribution
$\nu(\rho)$. Averaging over instanton positions in $d=4$ Euclidean
space--time produces the overall conservation of momenta of quarks
entering the vertex $Y$, hence it is convenient to write down
the quark interaction vertex in the momentum space. There are
formfactor functions $F(k\rho)$ \ur{Ff} associated with the Fourier
transform of the fermion zero modes of one instanton, attached to each
quark line entering the vertex. The $2N_f$-fermion vertex induced
by an instanton is, in momentum space,

\[
Y_{N_f}^+=\int\!d\rho\:\nu(\rho)\int\!
dU\prod_{f=1}^{N_f}\left\{\int\!\frac{d^4k_f}{(2\pi)^4}\: 2\pi\rho
F(k_f\rho) \int\!\frac{d^4l_f}{(2\pi)^4}\: 2\pi\rho
F(l_f\rho)\right.
\]
\beq
\left.\cdot(2\pi)^4\delta(k_1+...+k_{N_f}-l_1-...-l_{N_f})
\cdot U_{i^\prime_f}^{\alpha_f}U_{\beta_f}^{\dagger j^\prime_f}
\epsilon^{i_f i^\prime_f}\epsilon_{j_f j^\prime_f}
\left[i\psi_{Lf\alpha_fi_f}^\dagger(k_f)\psi_L^{f\beta_fj_f}(l_f)
\right]\right\};
\la{Y0}\eeq
for the $Y^-$ vertices induced by $\bar I$'s one has to replace
left-handed Weyl spinors $\psi_L, \psi^\dagger_L$ by
right-handed ones, $\psi_R, \psi^\dagger_R$.
Using these vertices one can write down the partition function
to which QCD is reduced at low momenta, as a functional integral
over quark fields \cite{DP4,D1,DPW}:

\beq
{\cal Z} =\int\!\!D\psi D\psi^\dagger
\exp\left(\int\!d^4x\sum_{f=1}^{N_f}\bar\psi_f i\dd \psi^f\right)
\left(\frac{Y_{N_f}^+}{VM_1^{N_f}}\right)^{N_+}
\left(\frac{Y_{N_f}^-}{VM_1^{N_f}}\right)^{N_-}
\la{Z0}\eeq
where $N_\pm$ are the number of \IIs in the whole $d=4$ volume
$V$. The volume factors in the denominators arise because of
averaging over individual instanton positions, and certain mass
factors $M_1^{N_f}$ are put in to make \eq{Z0} dimensionless.
Actually, the mass parameter $M_1$ plays the role of separating
high-frequency part of the fermion determinant in the instanton
background from the low-frequency part considered here.
Its concrete value is irrelevant for the derivation
of the low-energy effective action performed below; in fact it is
established from smooth matching of high- and low-frequency
contributions to the full fermion determinant in the instanton
vacuum \cite{DP2}.

Having fermion interactions in the pre-exponent of the partition
function is not convenient: one should rather have the
interactions in the exponent, together with the kinetic energy
term. This can be achieved by rewriting \eq{Z0} with the help
of additional integration over `Lagrange multipliers'
$\lambda_\pm$:

\[
{\cal Z} =
\int\!\frac{d\lambda_\pm}{2\pi}\int\!\!D\psi D\psi^\dagger
\exp\left\{N_+\left(\ln\frac{N_+}{\lambda_+VM_1^{N_f}}-1\right)
+N_-\left(\ln\frac{N_-}{\lambda_-VM_1^{N_f}}-1\right)\right.+
\]
\beq
\left.+\int\!d^4x
\sum_{f=1}^{N_f}\bar\psi_f i\dd \psi^f
+\lambda_+Y_{N_f}^++\lambda_-Y_{N_f}^-
\right\}\!.
\la{Z1}\eeq
Since $N_\pm\sim V\rightarrow\infty$ integration over
$\lambda_\pm$ can be performed by the saddle-point method; the
result is \eq{Z0} we started from.

As seen from \eq{Z1} $\lambda_\pm$ plays the role of the coupling
constant in the many-quark interactions. It is very important
that their strength is not pre-given but is, rather, determined
self-consistently from the fermion dynamics itself; in
particlular, the saddle-point values of $\lambda_\pm$ depend on
the phase quarks assume in the instanton vacuum. As shown below,
in the chiral symmetry broken phase the values of $\lambda_\pm$,
as determined by a saddle-point equation, appear to be real.

To get the $2N_f$-fermion vertices \ur{Y0} in a closed form one
has to explicitly integrate over instanton orientations in colour
space. For the $N_f$-fermion vertex one has to
average over $N_f$ pairs of $(U, U^\dagger)$. In particular,
one has:

\[
\int dU=1,\;\;\;\;\int dU\:U_i^\alpha U_\beta^{\dagger j}
=\frac{1}{N_c}\delta_\beta^\alpha\delta_i^j,\;\;\;\;\;
\int dU\:U_{i_1}^{\alpha_1}U_{i_2}^{\alpha_2}
U_{\beta_1}^{\dagger j_1}U_{\beta_2}^{\dagger j_2}
\]
\beq
=\frac{1}{N_c^2-1}\left[\delta^{\alpha_1}_{\beta_1}
\delta^{\alpha_2}_{\beta_2}
\left(\delta^{j_1}_{i_1}\delta^{j_2}_{i_2}-\frac{1}{N_c}
\delta^{j_1}_{i_2}\delta^{j_2}_{i_1}\right)
+\delta^{\alpha_1}_{\beta_2}\delta^{\alpha_2}_{\beta_1}
\left(\delta^{j_1}_{i_2}\delta^{j_2}_{i_1}-\frac{1}{N_c}
\delta^{j_1}_{i_1}\delta^{j_2}_{i_2}\right)\right],
\;\;\;{\rm etc.}
\la{aver}\eeq

We present below the resulting vertices for $N_f=1,2,3$ and
for any $N_f$ but $N_c\rightarrow\infty$.

\vskip .5true cm
\underline{${\bf N_f=1}$}
\vskip .5true cm

\noindent In this case the ``vertex'' \ur{Y0} is just a mass term
for quarks,

\beq
Y^\pm_1=\frac{i}{N_c}\int\frac{d^4k}{(2\pi)^4}
\int\!d\rho\:\nu(\rho)[2\pi\rho F(k\rho)]^2
\left[\psi_\alpha^\dagger(k)\frac{1\pm\gamma_5}{2}\psi^\alpha(k)
\right]\!,
\la{Y1}\eeq
with a momentum dependent dynamically-generated mass $M(k)$ given by

\beq
M(k)=\frac{\lambda}{N_c}\int\!d\rho\:\nu(\rho)\left[2\pi\rho
F(k\rho)\right]^2 \approx\frac{\lambda}{N_c}\left[2\pi\bar\rho
F(k\bar\rho)\right]^2.
\la{momdep}\eeq

In all our previous work on the instanton vacuum we have assumed
that the distribution in the sizes of instantons, $\nu(\rho)$,
is a sharp function peaked at certain $\bar\rho$, and replaced $\rho$
by this $\bar\rho$ in the argument of the formfactor functions
$F(k\rho)$. However, there is a subtlety here: if the size
distribution for large $\rho$ behaves as $\nu(\rho)\sim 1/\rho^3$
(corresponding to the linear potential between heavy quarks
\cite{DPP1}) the dynamical quark mass logarithmically diverges
at small momenta implying, in a sense, the confinement of light
quarks. We shall not pursue this interesting topic here but replace each
time $\rho$ by its average value $\bar\rho$.

In order to find the overall scale $\lambda$ of the dynamical mass
one has to put \ur{Y1} into \eq{Z1}, integrate over
fermions, and find the minimum of the free energy in respect to
$\lambda_\pm$.  At $\theta=0$ the QCD vacuum is $CP$ invariant so that
$N_+=N_-=N/2$ and consequently $\lambda_+=\lambda_-=\lambda$
\footnote{Fluctuations of the topological charge, $N_+-N_-$, leading to
the so-called topological susceptibility (related to the solution of
the $U(1)$ problem) has been considered in refs.\cite{DP4,DPW}.}.
In this case the $\gamma_5$ term in $Y^\pm$ gets cancelled, and the
exponent of the partition function \ur{Z1} reads:

\[
-N\ln \lambda+\int\!d^4x\int\!\frac{d^4k}{(2\pi)^4}
\Tr\ln\left\{\dk+i\frac{\lambda}{N_c}[2\pi\bar\rho
F(k\bar\rho)]^2 \right\}
\]
\beq =-N\ln \lambda+2N_cV\int\!\frac{d^4k}{(2\pi)^4}
\ln\left\{k^2+\left(\frac{\lambda}{N_c}[2\pi\bar\rho F(k\bar\rho)]^2
\right)^2\right\}.
\la{Z11}\eeq

Differentiating it in respect to $\lambda$ and using
\eq{momdep} one gets the gap \eq{selfcons} or the
self-consistency condition which is in fact a requirement on the
overall scale of the constituent quark mass $M(k)$; its momentum
dependence is anyhow given by \eq{momdep}.  Since the momentum
integration in \eq{selfcons} is well convergent and is actually
cut at momenta $k\sim 1/\bar\rho$, the saddle-point value of the
`Lagrange multiplier' $\lambda$ is of the order of
$\sqrt{N_cN/V}/\bar\rho$. The steepness of the saddle-point
integration is proportional to the volume $V$, hence the use
of the saddle-point method is absolutely justified.

Note that \eq{Y1} reproduces the massive quark propagator \ur{qprop},
hence the chiral condensate is given by \eq{chircond}. It is very
important that, initially, one does not know the strength of the quark
interactions (represented by the `Lagrange multiplier' $\lambda$): it
is fixed only after integration over the quark fields is performed. We
also stress that the basic quantities associated with spontaneous
chiral symmetry breaking, such as $<\bar\psi\psi>, M$ or $F_\pi$ are
{\em non-analytic} in the instanton density, $N/V$: such a behaviour is
characteristic of spontaneous breaking of continuous symmetry.

\vskip .5true cm
\underline{${\bf N_f=2}$}
\vskip .5true cm

In this case averaging \eq{Y1} over the instanton orientations with
the help of \eq{aver} gives a nontrivial 4-fermion interaction.
It is, of course, non-local: a formfactor function $F(k\rho)$ is
attributed to each fermion entering the vertex; in addition
it should be averaged over the sizes of instantons. The non-locality
is thus of the order of the average instanton size in the vacuum.
One has \cite{DP4,DP5}:

\[
Y_2^+=\frac{i^2}{N_c^2-1}\int\frac{d^4k_1d^4k_2d^4l_1d^4l_2}
{(2\pi)^{12}}\delta(k_1+k_2-l_1-l_2)
\]
\[
\cdot\int\!d\rho\:\nu(\rho)\:(2\pi\rho)^4
F(k_1\rho)F(k_2\rho)F(l_1\rho)F(l_2\rho)
\]
\[
\cdot\frac{1}{2!}\epsilon^{f_1f_2}\epsilon_{g_1g_2}\left\{
\left(1-\frac{1}{2N_c}\right)[\psi_{Lf_1}^\dagger(k_1)\psi_L^{g_1}(l_1)]
[\psi_{Lf_2}^\dagger(k_2)\psi_L^{g_2}(l_2)]\right.
\]
\beq
\left. +\frac{1}{8N_c}
[\psi_{Lf_1}^\dagger(k_1)\sigma_{\mu\nu}\psi_L^{g_1}(l_1)]
[\psi_{Lf_2}^\dagger(k_2)\sigma_{\mu\nu}\psi_L^{g_2}(l_2)]\right\}\!.
\la{Y2}\eeq
For the $\bar I$-induced vertex $Y^-$ one has to replace
left-handed components by right-handed ones. In all square
brackets summation over colour is understood. Note that the
last-line (tensor) term is suppressed at large $N_c$; it,
however, is crucial at $N_c=2$ to support the actual $SU(4)$
chiral symmetry in that case \cite{DP5}. The antisymmetric
$\epsilon^{f_1f_2}\epsilon_{g_1g_2}$ structure demonstrates that
the interactions have a determinant form in the two flavours.
Using the identity

\beq
2\epsilon^{f_1f_2}\epsilon_{g_1g_2}=\delta_{g_1}^{f_1}\delta_{g_2}^{f_2}
-(\tau^A)_{g_1}^{f_1}(\tau^A)_{g_2}^{f_2}
\la{id}\eeq
and adding the $\bar I$-induced vertex $Y_2^-$
one can rewrite the leading-$N_c$ (first) term of \eq{Y2} as

\beq
(\psi^\dagger\psi)^2+(\psi^\dagger\gamma_5\psi)^2
-(\psi^\dagger\tau^A\psi)^2-(\psi^\dagger\tau^A\gamma_5\psi)^2
\la{NJLf}\eeq
which resembles closely the Vaks--Larkin \cite{VL} /
Nambu--Jona-Lasinio \cite{NJL} model. It should be stressed
though that in contrast to that {\em at hoc} model the
interaction \ur{Y2} {\it i}) violates explicitly the $U_A(1)$
symmetry, {\it ii}) has a fixed interaction strength related
to the density of instantons (see below) and {\it iii}) contains
an intrinsic ultraviolet cutoff due to the formfactor functions
$F(k\rho)$. In addition, at $N_c=2$ it correctly preserves
the actual $SU(4)$ chiral symmetry. We shall show in the next
section that the four-fermion interaction \ur{Y2} leads to the
spontaneous chiral symmetry breaking, with the appearance of the
constituent quark mass $M(k)$ satisfying the same gap equation
\ur{selfcons} as we obtained above in the case $N_f=1$.

\vskip .5true cm
\underline{${\bf N_f=3}$}
\vskip .5true cm

In this case one gets a 6-fermion vertex of the following structure
\cite{DP5}:

\[
Y_3^+=\frac{i^3}{N_c(N_c^2-1)}
\int\frac{d^4k_1d^4k_2d^4k_3d^4l_1d^4l_2d^4l_3}
{(2\pi)^{20}}\delta(k_1+k_2+k_3-l_1-l_2-l_3)
\]
\[
\cdot\int\!d\rho\:\nu(\rho)\:(2\pi\rho)^6
F(k_1\rho)F(k_2\rho)F(k_3\rho)F(l_1\rho)F(l_2\rho)F(l_3\rho)
\]
\[
\cdot\frac{1}{3!}\epsilon^{f_1f_2f_3}\epsilon_{g_1g_2g_3}
\left\{\left(1-\frac{3}{2(N_c+2)}\right)
[\psi_{Lf_1}^\dagger(k_1)\psi_L^{g_1}(l_1)]
[\psi_{Lf_2}^\dagger(k_2)\psi_L^{g_2}(l_2)]
[\psi_{Lf_3}^\dagger(k_3)\psi_L^{g_3}(l_3)]
\right.
\]
\beq
\left. +\frac{3}{8(N_c+2)}
[\psi_{Lf_1}^\dagger(k_1)\psi_L^{g_1}(l_1)]
[\psi_{Lf_2}^\dagger(k_2)\sigma_{\mu\nu}\psi_L^{g_2}(l_2)]
[\psi_{Lf_3}^\dagger(k_3)\sigma_{\mu\nu}\psi_L^{g_3}(l_3)]\right\}\!.
\la{Y3}\eeq
Again, it is of a determinant structure, this time in 3 flavours,
and again the tensor term is suppressed at $N_c\rightarrow\infty$.

\begin{figure}
\centerline{\epsfxsize10.0cm\epsffile{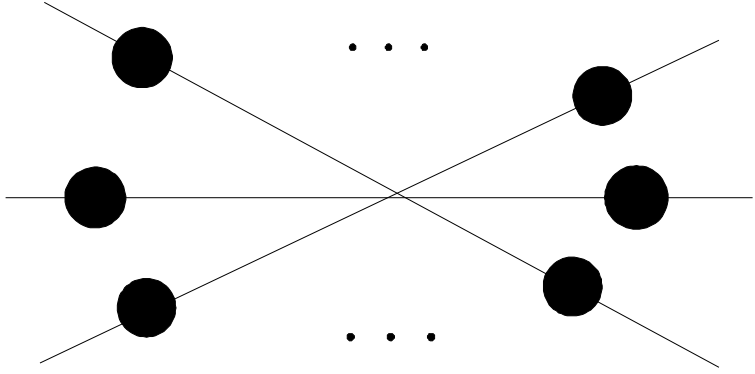}}
\caption[]{Instanton-induced $2N_f$-quark vertex. The black
blobs denote the formfactor functions $F(k\rho)$ attached to each
quark leg.}
\end{figure}

\vskip .5true cm
\underline{{\bf Any} ${\bf N_f}$}
\vskip .5true cm

For arbitrary $N_f$ the {\em leading} term at $N_c\rightarrow\infty$
can be written as a determinant of $N_f\times N_f$ matrices composed
of non-local chiral quark bilinears $J_{fg}$ \cite{DP4,DP5,DPW}:

\beq
Y_{N_f}^\pm\stackrel{N_c\rightarrow\infty}{=}
\left(\frac{1}{N_c}\right)^{N_f}
\int\!d^4x\int\!d\rho\:\nu(\rho)\;
{\rm det}_{N_f}\left[iJ^\pm(x,\rho)\right]
\la{YNf}\eeq
where

\beq
J_{fg}^\pm(x,\rho)=\int\frac{d^4kd^4l}{(2\pi)^8}e^{i(k-l,x)}
[2\pi\rho F(k\rho)][2\pi\rho F(l\rho)]
\left[\psi_{f\alpha}^\dagger(k)\frac{1\pm\gamma_5}{2}\psi^{g\alpha}(l)
\right].
\la{Jfg}\eeq
If one neglects formfactors, that is puts $F=1$, these chiral
currents become local,

\beq
J_{fg}^\pm(x,\rho)\approx(2\pi\rho)^2
\left[\psi_{f\alpha}^\dagger(x)\frac{1\pm\gamma_5}{2}\psi^{g\alpha}(x)
\right].
\la{Jfgloc}\eeq

\subsection{Bosonization}

One can linearize the many-fermion vertices induced by instantons
by introducing auxiliary boson fields. This formal procedure is
called bosonization of the theory. Roughly speaking, when one
has a theory with 4-fermion interactions, it can be viewed as a
limit of a one-boson exchange when the mass of intermediate boson
tends to infinity. This is the meaning of the bosonization.

In case $N_f=2$ the instanton-induced interactions are 4-fermion ones
(see \eqs{Y2}{NJLf}) and it is very easy to bosonize them by
introducing scalar and pseudoscalar fields. Note that the non-leading
second term in \eq{Y2} requires additional tensor fields for the
bosonization. In case $N_f\geq 3$ the instanton-induced interactions
become 6-fermion and so forth, so that the bosonization becomes
less trivial. However, it is still possible to perform it using
$N_c$ as an algebraically large parameter \cite{DP4,DP5,DPW}.
Indeed, introducing $N_f\times N_f$ matrices ${\cal M}$ the
following equation becomes true in the saddle-point approximation
(justified at large $N_c$):

\beq
\exp\left(\lambda {\rm det}\left[\frac{iJ}{N_c}\right]\right)
=\int\!d{\cal M}
\exp\left\{i{\rm Tr}({\cal M}J)-(N_f-1)
\left(\frac{{\rm det}
[{\cal M}N_c]}{\lambda}\right)^{\frac{1}{N_f-1}}\right\}\!.
\la{sadpoi}\eeq

This remarkable formula enables one to bosonize many-fermion
interactions of the determinant type, like in \eq{YNf}. It should
be stressed however that the procedure is justified only at large
$N_c$, otherwise {\em i)} the fermion interactions have not the
simple determinant form and {\em ii)} the saddle-point evaluation
of the integral in \eq{sadpoi} is not justified. We notice that
at $N_f=2$ \eq{sadpoi} becomes exact since in this case the
integral over ${\cal M}$ is Gaussian. Indeed, in this particular
case the power of $\det{\cal M}$ is unity while the determinant
of a $2\times 2$ matrix itself is quadratic in matrix entries.

We are now prepared to write down the partition function \ur{Z1}
by introducing auxiliary integration over `meson' fields
${\cal M}_{L,R}(x)$ coupled linearly to the quark chiral currents
$J^\pm(x)$; both quantities are $N_f\times N_f$ matrices in
flavour. We have \cite{DP5,DPW}:

\[
{\cal Z}=\int\!\frac{d\lambda}{2\pi}\:\exp(-N\ln\lambda)
\]
\[
\cdot\int\!D{\cal M}_{L,R}\:\exp\int\!d^4x\left\{-(N_f-1)\left[
\left(\frac{{\rm det}[{\cal M}_LN_c]}{\lambda}\right)^{\frac{1}{N_f-1}}
+\left(\frac{{\rm det}[{\cal M}_RN_c]}{\lambda}\right)^{\frac{1}{N_f-1}}
\right]\right\}
\]
\beq
\cdot \int\!D\psi^\dagger D\psi\:\exp\int\!d^4x
\left\{\psi_f^\dagger i\dd \psi^f
+i\Tr[{\cal M}_LJ^+]+i\Tr[{\cal M}_RJ^-]\right\}\!.
\la{ZNf}\eeq
The last line presents a theory of quarks interacting with external
chiral `meson' fields, ${\cal M}_{L,R}$. Integration over quarks
can be performed by expanding the resulting functional fermion
determinant in powers of ${\cal M}_{L,R}$ and / or their derivatives.
A concrete example of such expansion will be given below.

The second line in \eq{ZNf} presents the potential energy of
the ${\cal M}_{L,R}$ fields: it has a rather peculiar form of a power
of the determinant composed of these fields. In the particular case
$N_f=1$ these terms vanish, as it should be since at $N_f=1$
instantons induce just a mass term for fermions, see the previous
subsection. In the particular case of $N_f=2$ the potential energy of
the ${\cal M}_{L,R}$ fields becomes quadratic in the fields, that
is to say, we have just a mass term for the `meson' fields. Notice
that there is no kinetic energy term at the tree level: the kinetic
energy (as well as higher derivative terms) is generated dynamically
after one integrates over quarks, that is through quark loops.

The fermion action (last line in \eq{ZNf}) is invariant under
full chiral rotations with arbitrary $U(N_f)$ matrices $A,B$:

\beq
\left\{\begin{array}{ccc}
\psi_L\rightarrow A\psi_L,\;\;&\;\;
\psi_L^\dagger\rightarrow \psi_L^\dagger B^\dagger,\;\;&\;\;
{\cal M}_L\rightarrow B{\cal M}_L A^\dagger,\\
\psi_R\rightarrow B\psi_R,\;\;&\;\;
\psi_R^\dagger\rightarrow \psi_R^\dagger A^\dagger,\;\;&\;\;
{\cal M}_R\rightarrow A{\cal M}_R B^\dagger.
\end{array}\right.
\la{unirot}\eeq
However, the potential energy of the `meson' fields (the second line
in \eq{ZNf}) has a smaller invariance. Indeed the determinants
transform as

\beq
\det[{\cal M}_L]\rightarrow \det[A^\dagger B]\det[{\cal M}_L],
\;\;\;\;\;
\det[{\cal M}_R]\rightarrow \det[B^\dagger A]\det[{\cal M}_R],
\la{dettrans}\eeq
therefore they acquire a $U(1)$ phase factor of the relative
$A^\dagger B$ transformation. For that reason \eq{ZNf} breakes
{\em explicitly} the axial $U_A(1)$ symmetry, as expected on general
grounds from instantons \cite{tH}.

Let us show that \eq{ZNf} leads to the {\em spontaneous} breaking
of the chiral symmetry. To that end, let us parametrize the `meson'
fields by

\beq
\begin{array}{c}
{\cal M}_L(x)=[\sigma(x)+\eta(x)]U(x)V(x),\\
{\cal M}_R(x)=[\sigma(x)-\eta(x)]V(x)U^\dagger(x)
\end{array}
\la{param1}\eeq
where $\sigma(x)$ is the scalar flavour-singlet field and $\eta(x)$
is the pseudoscalar flavour-singlet field. The $SU(N_f)$ matrix fields
$U(x)$ and $V(x)$ can be further on parametrized by scalar
$\sigma^A(x)$ and pseudoscalar $\pi^A(x)$ fields belonging to the
adjoint representation of the flavour $SU(N_f)$ group:

\beq
\begin{array}{c}
U(x)=\exp\:i\pi^A(x)\lambda^A,\\
V(x)=\exp\:i\sigma^A(x)\lambda^A
\end{array}
\la{param2}\eeq
where $\lambda^A$ are the Hermitean generators of the $SU(N_f)$.

The vacuum state of the theory given by the partition function \ur{ZNf}
corresponds to a non-zero value of the flavour-singlet $\sigma$ field.
This is equivalent to the spontaneous breakdown of chiral symmetry, and
the $\pi^A$ fields become Goldstone particles.

To reveal the nonzero value of the $\sigma$ field let us calculate
the effective potential for it. Putting

\beq
\sigma=const,\;\;\;\;\eta=0.\;\;\;\;\;U=V=1
\la{param3}\eeq
and integrating over quarks in the constant background scalar field
we obtain the effective potential $V_{eff}(\sigma,\lambda)$
to which we add the `Lagrange multiplier' piece, $N\ln\lambda$
(see \eq{ZNf}):

\beq
V_{eff}(\sigma,\lambda)=\frac{N}{V}\ln\lambda
+2(N_f-1)(\sigma N_c)^{\frac{N_f}{N_f-1}}
\lambda^{-\frac{1}{N_f-1}}-2N_fN_c\int\!\frac{d^4k}{(2\pi)^4}
\:\ln\left\{k^2+\sigma^2\left[2\pi\bar\rho F(k\bar\rho)\right]^4
\right\}\!.
\la{effpot}\eeq
We have now to minimize \ur{effpot} in both quantities, $\sigma$ and
$\lambda$. The extremum condition gives the following
equations for the saddle-point values $\sigma_0, \lambda_0$:

\beq
\frac{N}{V}=4N_c\int\!\frac{d^4k}{(2\pi)^4}\:\frac{M^2(k)}{k^2+M^2(k)},
\;\;\;\;\;M(k)=\sigma_0\left[2\pi\bar\rho F(k\bar\rho)\right]^2,
\la{1st}\eeq
\beq
\lambda_0=\frac{N}{2V}\left(\frac{2\sigma_0\:VN_c}{N}\right)^{N_f}.
\la{2nd}\eeq

These equations demonstrate that the scalar field develops a
nonzero v.e.v. $\sigma_0$, quarks get a dynamical mass $M(k)$
and chiral symmetry is broken.
The first equation is the familiar self-consistency or gap equation
\ur{selfcons} which we have also obtained in the $N_f=1$ case: it
determines the overall scale of the dynamically generated
momentum-dependent constituent quark mass $M(k)$. The second \eq{2nd}
relates the value of the `Lagrange multiplier' $\lambda$ to the
v.e.v. of the $\sigma$ field. We find that these quantities
have the following parametric dependence on the basic characteristics
of the instanton vacuum, their density, $N/V$, and their average
size, $\bar\rho$:

\beq
M(0)\approx\sigma_0(2\pi\bar\rho)^2
\sim\sqrt{\frac{N\pi^2\bar\rho^2}{VN_c}},\;\;\;\;\;
\lambda_0\sim\frac{N}{V}\left(\frac{VN_c}{N\pi^2\bar\rho^2}\right)^
{\frac{N_f}{2}}.
\la{parametric}\eeq

It is important that the strength $\lambda$ of the $2N_f$-fermion
instanton-induced interactions is not fixed beforehand but is, rather,
determined by the phase the fermion system assumes. In a given phase,
like the chirality broken phase, the coupling constant $\lambda$ is
defined unambigiously through the extremum conditions \urs{1st}{2nd}.
In another phase, say, chiral invariant or in a phase where diquarks
condense \cite{DP5}, the saddle-point value of $\lambda$ would be,
generally speaking, different. It means that the formalism presented
here is different from the `quenched approximation': it incorporates
the back reaction of fermions on the instanton medium. The main
assumption in the starting formula for this derivation, \eq{Z0}, is
that one can average independently over the collective coordinates of
\IIs. It implies that correlations between pseudoparticles coming from
the gauge sector are neglected or, better to say, treated {\em \`a la}
variational principle resulting in an effective size distribution
$\nu(\rho)$ \cite{DP1,DPW}, however correlations arising from fermions
are taken into account.

In getting the equation for the v.e.v. of the scalar field $\sigma$
we have used the mean-field approximation, \eq{1st}. Theoretically
speaking, its accuracy is of the order of $O(N_f/N_c)$ since the
meson loops have been disregarded. In principle, they could be taken
into account. It should be reminded, though, that \eq{ZNf}
itself has been derived in the limit of $N_c\rightarrow\infty$;
therefore, if one wishes to take into account higher order
corrections in $N_f/N_c$ one should rather work with the unabridged
vertices \ur{Y2} or \ur{Y3}.

\subsection{Chiral lagrangian}

The low-momentum QCD partition function \ur{ZNf}, after integration
over quarks in the given background `meson' fields
${\cal M}_{L,R}$, gives an effective action for $N_f^2$ scalar
and $N_f^2$ pseudoscalar fields which one can parametrize according
to \eqs{param1}{param2}. The scalar flavour-singlet field $\sigma(x)$
should be counted from its mean-field value $\sigma_0$
given by \eq{1st}, and the `Lagrange multiplier' $\lambda$ should be
put to its saddle-point value \ur{2nd}.

The $2N_f^2$ fields introduced by \eqs{param1}{param2} are not properly
normalized: to get the correct normalization one has to extract
their kinetic energy (= two derivative) terms from the quark functional
determinant and redifine the fields so that the kinetic energy is
standard, $\frac{1}{2}(\partial_\mu\phi)^2$. The cubic, quartic,...
terms in the boson fields arising from \eq{ZNf} will then give the
meson interactions. It can be shown from $N_c$ power counting
\cite{DP4} that the cubic coupling constants for the properly normalized
meson fields are of the order of $1/\surd N_c$, the quartic couplings
are $\sim 1/N_c$, and so on. This is as it should be from the general
$N_c$ counting rules.

We have already explained in the previous subsection that the axial
$U_A(1)$ symmetry is explicitly (not spontaneously!) broken in
\eq{ZNf}, therefore the pseudoscalar flavour-singlet $\eta$ meson
is not a Goldstone boson, the $U_A(1)$ problem is solved. Moreover,
in the limit $N_f/N_c\rightarrow 0$ we recover
from \eq{ZNf} \cite{DP4,DPW}  the theoretical Witten--Veneziano formula
for the singlet $\eta^\prime$ mass, as given by

\beq
m_{\eta^\prime}^2=\frac{2N_f<Q_T^2>/V}{F_\pi^2}
\la{etap}\eeq
where $<Q_T^2>/V=<(N_+-N_-)^2>/V$ is the topological susceptibility of
the vacuum.

As to the non-singlet pseudoscalar fields $\pi^A(x)$ introduced
by \eqs{param1}{param2} which we shall call pions for short, they
appear to be massless Goldstone fields. Indeed, the constant fields
$\pi^A$ correspond to global chiral rotations \ur{unirot} with
$A^\dagger=B=\exp (i\pi^A\lambda^A)$. \Eq{ZNf} is invariant under
such rotations, therefore the lagrangian for the $\pi^A(x)$ fields
contain only derivatives, {\em i.e.} pions are massless -- in
accordance with the Goldstone theorem.

One can check from \eq{ZNf} \cite{DP4} that scalar meson fields
get the mass of the order of $1/\bar\rho$; numerically it is in the
$1\:GeV$ range. Though the $\eta^\prime$ mass, algebraically, is
given by a different formula \ur{etap}, numerically it also turns
out to be about $1\:GeV$. Strictly speaking, in that range of
momenta the low-energy QCD partition function \ur{ZNf} is not
justifiable as it has been derived above from the partition
function \ur{Z0} valid for momenta $k\le 1/\bar\rho$. Therefore,
in a consistent approach to the low-momentum theory at
$k\le 1/\bar\rho$ one has to freeze out all the meson fields except
massless pions. [In the academic limit of $N_f/N_c\rightarrow 0$
one would also need to keep the $\eta^\prime$ degree of freedom.]
It is very important that the quark masses are, parametrically
speaking, much less than $1/\bar\rho$: the dimensionless
quantity

\beq
\left(M\bar\rho\right)^2\sim\frac{1}{N_c}
\pi^2\bar\rho^4\frac{N}{V}\;\ll\;1
\la{p}\eeq
is suppressed by the packing fraction of instantons in the vacuum.
The whole approach to the instanton vacuum implies that instantons
are on the average relatively dilute and that this packing fraction 
is numerically small. Theoretically, the smallness of the parameter 
\ur{p} can be traced back to the ``accidentally'' large coefficient in 
the Gell-Mann--Low function, the famous $11/3$ \cite{DP1,DPW}.

We, thus, arrive to the conclusion that at low momenta $k\le
1/\bar\rho$ there are exactly two degrees of freedom left:
quarks with a dynamical mass $M\ll 1/\bar\rho$ and the massless
Goldstone pions. In principle, one could think of a `soft'
instanton size distribution going at large sizes as
$\nu(\rho)\sim 1/\rho^3$. Such a distribution is peculiar
because it automatically leads to a linear potential between
heavy quarks \cite{DPP1} and could therefore reproduce confinement.
Moreover, there are certain reasons to believe that this
particular distribution is, effectively, realized in nature
[Diakonov and Petrov, in preparation]. As explained above
in such a case $M(k)$ would logarithmically diverge at
$k\rightarrow 0$ (see \eq{momdep}) so that free quarks would
not show up. Amusingly, such a possibility would not invalidate the use
of the effective chiral theory for {\em bound state} problems, like
inside the nucleons, since for virtualities $k\sim ({\rm
nucleon~radius})^{-1}$ the estimate \ur{p} would still hold true. In
these lectures, however, we shall not pursue this interesting
possibility of marrying confinement with the chiral theory but assume
that averaging over instanton sizes merely replaces $\rho$ by its peak
or average value $\bar\rho$.

Having made these preliminary remarks, let us write down the
effective partition function to which QCD is reduced at low momenta,
$k\le 1/\bar\rho$. It follows from the instanton-induced partition
function \ur{ZNf} where we freeze out all meson fields except pions
and put $\sigma$ and $\lambda$ to their saddle-point values:

\[
{\cal Z}=\int\!D\pi^A\int\!D\psi^\dagger D\psi
\exp\int\!d^4x \left\{\psi_f^\dagger(x) i\dd \psi^f(x)
+i\int\!\frac{d^4kd^4l}{(2\pi)^8}e^{i(k-l,x)}\sqrt{M(k)M(l)}
\right.
\]
\beq
\left.
\cdot\left[\psi_{f\alpha}^\dagger(k)\left(U^f_g(x)
\frac{1+\gamma_5}{2}+U^{\dagger f}_g(x)\frac{1-\gamma_5}{2}
\right)\psi^{g\alpha}(l)\right]\right\},\;\;\;\;
U^f_g(x)=\left(\exp i\pi^A(x)\lambda^A)\right)^f_g.
\la{Znash}\eeq
This effective theory has been first derived in refs. \cite{DP3,DP4}.
\Eq{Znash} shows quarks interacting with chiral fields $U(x)$, with
formfactor functions equal to the square root of the dynamical quark
mass attributed to each vertex where $U(x)$ applies. The matrix
entering in the parentheses is actually a $N_f\times N_f$ matrix in
flavour and a $4\times 4$ matrix in Dirac indices. It can be
identically rewritten as

\beq
U(x)\frac{1+\gamma_5}{2}
+U^\dagger (x)\frac{1-\gamma_5}{2}
=\exp\left(i\pi^A(x)\lambda^A\gamma_5\right)\equiv U^{\gamma_5}(x),
\la{pasha}\eeq
the industrious final abbreviation being due to Pavel Pobylitsa.

The formfactor functions $\surd M(k\bar\rho)$ for each quark
line attached to the chiral vertex automatically cut off momenta
at $k\ge 1/\bar\rho$. In the range of quark momenta $k\ll 1/\bar\rho$
(which we shall be mostly interested in) one can neglect this
non-locality, and the partition function \ur{Znash} is simplified
to a local field theory:

\beq
{\cal Z}=\int\!D\pi^A\int\!D\psi^\dagger D\psi\;
\exp\int\!d^4x\: \psi^\dagger(x)\left[i\dd+iMU^{\gamma_5}(x)\right]
\psi(x).
\la{Zna}\eeq
One should remember, however, to cut the quark loop integrals
at $k\approx 1/\bar\rho\approx 600\:MeV$. Notice that there is
no kinetic energy term for pions: it appears only after one integrates
over the quark loop, see below. Summation over colour is assumed in the
exponent of \eq{Zna}.

\Eq{Zna} defines a simple and elegant local field theory though it is
still a highly non-trivial one. Its main properties will be established
in the next section \footnote{We have been asked about the relation
of this low-energy theory with that suggested by Manohar and Georgi
\cite{MG}. One can redefine the quark fields

\beq
\psi\rightarrow \psi^\prime=\exp(i\pi^A\lambda^A\gamma_5/2)\psi,
\;\;\;\;\;
\psi^\dagger\rightarrow \psi^{\dagger\prime}=\psi^\dagger
\exp(i\pi^A\lambda^A\gamma_5/2),
\la{changevar}\eeq
and rewrite the lagrangian in \ur{Zna} as

\beq
{\cal L}=\psi^{\dagger\prime}(i\dd+\Dirac V + \Dirac A\gamma_5
+iM)\psi^\prime
\la{newecl}\eeq
with
\beq
V_\mu=\frac{i}{2}(\xi\partial_\mu\xi^\dagger
+\xi^\dagger\partial_\mu\xi),\;\;\;\;
A_\mu=\frac{i}{2}(\xi\partial_\mu\xi^\dagger
-\xi^\dagger\partial_\mu\xi),\;\;\;\;
\xi=\exp(i\pi^A\lambda^A/2)=U^{1/2},
\la{defs}\eeq
which resembles closely the effective lagrangian of Manohar and Georgi
(the effective chiral lagrangian in a similar form has been
independently suggested in ref. \cite{DE}).

The crucial difference is that Manohar and Georgi have added an
explicit kinetic energy term
$F_\pi^2\Tr(\partial_\mu U^\dagger \partial_\mu U)/4$
on top of \eq{newecl}. This is a typical double counting as the
kinetic energy term arises from quark loops, see the next section.}.

\vskip 1true cm
\section{Properties of the effective chiral lagrangian (EChL)}
\setcounter{equation}{0}

Properties of effective theories of quarks interacting with various
meson fields have been studied by several authors in the 80's, most
notably by Volkov and Ebert \cite{Volkov} and Dhar, Shankar and Wadia
\cite{Dhar}. The fact that integrating over quarks one gets, in
particular, the so-called Wess--Zumino term has been first established
by Eides and myself \cite{DE}, though in somewhat different settings,
see also below.

Integrating over the quark fields in \eq{Zna} one gets the
effective chiral lagrangian (EChL):

\beq
S_{eff}[\pi]=-N_c\ln\det\left(i\dd+iMU^{\gamma_5}\right).
\la{Seff}\eeq
The Dirac operator entering \eq{Seff} is not hermitean:

\beq
D=i\dd+iMU^{\gamma_5},\;\;\;\;\;D^\dagger=i\dd-iMU^{\gamma_5\dagger},
\la{herm}\eeq
therefore the effective action has an imaginary part. The real part
can be defined as

\[
{\rm Re} S_{eff}[\pi]=-\frac{N_c}{2}\ln\det\left(\frac{D^\dagger D}
{D_0^\dagger D_0}\right),
\]
\beq
D^\dagger D=-\partial^2+M^2-M(\dd U^{\gamma_5}),\;\;\;\;\;
D_0^\dagger D_0=-\partial^2+M^2.
\la{ReS}\eeq

In the next two subsections we establish the properties of the
real and imaginary parts of the EChL separately, following ref.
\cite{DPP2}.

\subsection{Derivative expansion and interpolation formula}

There is no general expression for the functional \ur{Seff} for
arbitrary pion fields. For certain pion fields the functional
determinant \ur{Seff} can be estimated numerically, see section 4.
However, one can make a systematic expansion of the EChL in increasing
powers of the derivatives of the pion field, $\partial U$. It is called
long wave-length or derivative expansion. Moreover, one can do even
better and expand the real part of the EChL in powers of

\beq
\frac{pM}{p^2+M^2}\:(U-1)
\la{interpopar}\eeq
where $p$ is the characteristic momentum of the pion field.
This quantity becomes small in three limiting cases: {\em i}) small
pion fields, $\pi^A(x)\ll 1$, with arbitrary momenta, {\em ii})
arbitrary pion fields but with small gradients or momenta, $p\ll M$,
{\em iii}) arbitrary pion fields and large momenta, $p\gg M$. We see
thus that expanding the EChL in this parameter one gets accurate
results in three corners of the Hilbert space of pion fields. For that
reason we call it {\em interpolation} formula \cite{DPP2}. Our
experience is that its numerical accuracy is quite good for more or
less arbitrary pion fields, even if one uses only the first term of
the expansion in \ur{interpopar}, see below.

The starting point for both expansions is the following formal
manipulation with the real part of the EChL \ur{ReS}.
The first move is to use the well-known formula, $\ln\det [{\rm
operator}] = \Sp\ln [{\rm operator}]$, where $\Sp$ denotes a
functional trace. One can write:

\[
{\rm Re} S_{eff}[\pi]=
-\frac{N_c}{2}\ln\det\left[1
-(-\partial^2+M^2)^{-1}M(\dd U^{\gamma_5})\right]
\]
\[
=-\frac{N_c}{2}\Sp\ln\left[1
-(-\partial^2+M^2)^{-1}M(\dd U^{\gamma_5})\right]
\]
\[
=-\frac{N_c}{2} \int\!d^4x\int\!\frac{d^4k}{(2\pi)^4}\:
e^{-ik\cdot x}\Tr\ln\left[1-(-\partial^2+M^2)^{-1}M
(\dd U^{\gamma_5})\right]e^{ik\cdot x}
\]
\beq
=-\frac{N_c}{2} \int\!d^4x\int\!\frac{d^4k}{(2\pi)^4}\:
\Tr\ln\left[1-(k^2+M^2-2ik\cdot\partial-\partial^2)^{-1}
M(\dd U^{\gamma_5})\right]\cdot 1,
\la{splog}\eeq
In going from the second to the third line we have written down
explicitly what does the functional trace $\Sp$ mean: take
matrix elements of the operator involved in a complete basis
(here: plane waves, $\exp (ik\cdot x)$), sum over all states (here:
integrate over $d^4k/(2\pi)^4$) and take the trace in $x$. `$\Tr$'
stands for taking not a functional but a usual matrix trace, in our
case both in flavour and Dirac bispinor indices. In going from the
third to the last line we have dragged the factor $\exp (ik\cdot x)$
through the operator, thus shifting all differential operators
$\partial\rightarrow\partial+ik$. We have put a unity at the end
of the equation to stress that the operator is acting on unity, in
particular, it does not differentiate it. The above is a standard
procedure for dealing with functional determinants \cite{DPY}.

The last expression in \eq{splog} can be now easily expanded in powers
of the derivatives of the pion field: it arises from expanding
\ur{splog} in powers of $\dd U^{\gamma_5}$ and of
$2ik\cdot\partial+\partial^2$. The first non-zero term has two
derivatives,

\[
{\rm Re} S_{eff}^{(2)}[\pi]=\frac{N_c}{4}
\int\!d^4x\int\!\frac{d^4k}{(2\pi)^4}\:\Tr\left(\frac{M\dd U^{\gamma_5}}
{k^2+M^2}\right)^2
\]
\beq
=\frac{1}{4}\int\!d^4x \Tr \left(\partial_\mu U^\dagger\partial_\mu U
\right)
\cdot 4N_c \int\!\frac{d^4k}{(2\pi)^4}\:\frac{M^2}{(k^2+M^2)^2}.
\la{S2}\eeq

It is the kinetic energy term for the pion field or, better to say,
the Weinberg chiral lagrangian: atually it contains all powers of the
pion field if one substitutes $U(x)=\exp(i\pi^A(x)\lambda^A)$. The
proportionality coefficient (the last factor in \eq{S2}) is called
$F_\pi^2$, experimentally, $F_\pi\approx 94\:MeV$. The last factor in
\eq{S2} is logarithmically divergent; to make it meaningful we have to
recall that we have actually simplified the theory when writing it in
the local form \ur{Zna}. Actually, the dynamical quark mass $M$ is
momentum-dependent (see \eq{Znash}); it cuts the logarithimically
divergent integral at $k\approx 1/\bar\rho$. Using the numerical
values of $\bar\rho\approx 600\:MeV$ and $M\approx 350\:MeV$ we
find

\beq
F_\pi^2=4N_c \int\!\frac{d^4k}{(2\pi)^4}\:\frac{M^2}{(k^2+M^2)^2}
\approx \frac{N_c}{2\pi^2}M^2\ln\frac{1}{M\bar\rho}
\approx (100\:MeV)^2
\la{Fpi1}\eeq
being not in a bad approximation to the experimental value of $F_\pi$.
Actually, the two-derivative term is the only divergent quantity in
the EChL: higher derivative terms are all finite.

A more standard way to present the two-derivative term is by using
hermitean $N_f\times N_f$ matrices $L_\mu=iU^\dagger\partial_\mu U$.
One can rewrite \eq{S2} as

\beq
{\rm Re} S_{eff}^{(2)}[\pi]
=\frac{F_\pi^2}{4}\int\!d^4x \Tr L_\mu L_\mu,\;\;\;\;\;
L_\mu=iU^\dagger\partial_\mu U.
\la{S21}\eeq

The next, four-derivative term in the expansion of ${\rm Re}S_{eff}$
is (note that the metric is Euclidean)

\beq
{\rm Re} S_{eff}^{(4)}[\pi]=-\frac{N_c}{192\pi^2}
\int\!d^4x\left[2\Tr(\partial_\mu L_\mu)^2+\Tr L_\mu L_\nu L_\mu L_\nu
\right].
\la{S4}\eeq

These terms describe, in particular, the $d$-wave $\pi\pi$ scattering
lengths, and other observables. They can be compared with the
appropriate phenomenological terms in the Gasser--Leutwyler chiral
perturbation theory \cite{GL}: \eq{S4} with the concrete
coefficients like $N_c/192\pi^2$ appears to be in good agreement
with the phenomenological analysis \cite{BCG}.
The next-to-next-to-leading six derivative terms following from
\eq{splog} have been computed in refs. \cite{Z,PV}; however, a detailed
comparison with phenomenology is still lacking here.

I would like to mention an interesting paper \cite{PolVer} where
the derivative expansion of the EChL has been obtained from dual
resonance models. For reasons not fully appreciated the dual
resonance model for the $\pi\pi$ scattering gives (numerically)
very similar coefficients as those following from \eq{S4}, however
there is a discrepancy at the 6-derivative level.

I now turn to the interpolation formula promised in the beginning of
this subsection. One can start from the last line in \eq{splog} and
expand in powers of $M(\dd\Ug)$. It is clear that the actual expansion
parameter will be \ur{interpopar}. In the first non-zero order
we get \cite{DPP2}

\beq
{\rm Re} S_{eff}^{interpol}[\pi]=\frac{N_c}{4}\int\!d^4x\int\!
\frac{d^4k}{(2\pi)^4}\:\Tr\left[\frac{1}{(k+i\partial)^2+M^2}
M(\dd U^{\gamma_5})\frac{1}{(k+i\partial)^2+M^2}M(\dd U^{\gamma_5})
\right].
\la{1interpol}\eeq
It will be convenient now to pass to the Fourier transform of the
$U(x)$ field understood as a matrix,

\beq
U(p)=\int\!d^4x\:e^{ip\cdot x}\left[U(x)-1\right].
\la{FTU}\eeq
The partial derivatives appearing in \eq{1interpol} act on the exponents
of the Fourier transforms of $U,U^\dagger$ and become corresponding
momenta. As a result we get

\beq
{\rm Re} S_{eff}^{interpol}[\pi]=\frac{1}{4}\int\!\frac{d^4p}{(2\pi)^4}
\:p^2\Tr\left[U^\dagger(p) U(p)\right]
\cdot 4N_c\int\!\frac{d^4k}{(2\pi)^4}
\frac{M^2}{\left[(k-\frac{p}{2})^2+M^2\right]
\left[(k+\frac{p}{2})^2+M^2\right]}\:.
\la{2interpol}\eeq
At $p\rightarrow 0$ the last factor becomes $F_\pi^2$ (see \eq{Fpi1}),
and \eq{2interpol} in nothing but the first term in the derivative
expansion, \eq{S2}. However, \eq{2interpol} also describes correctly
the functional $S_{eff}[\pi]$ for rapidly varying pion fields
(with momenta $p\gg M$) and for small pion fields of any momenta,
when one can anyhow expand \eq{splog} in terms of $\pi^A(x)$ and hence
in $U(x)-1$. The logarithmically divergent loop integral in
\eq{2interpol} should be regularized, as in \eq{Fpi1}.

Similarly, one can get the next term in the `interpolation' expansion
which will be quartic in $U(p)$, however our experience tells us that
already \eq{2interpol} gives a good approximation to the EChL for
most pion fields.

\subsection{The Wess--Zumino term and the baryon number}

We now consider the imaginary part of $S_{eff}[\pi]$. The first
non-zero term in the derivative expansion of ${\rm Im} S_{eff}[\pi]$
is \cite{DE,Dhar} the Wess-Zumino term \cite{WZ}. It is known \cite{W}
that it cannot be written as a $d=4$ integral over a local expression
made of the unitary $U(x)$ matrices, however the variation of the
Wess--Zumino term is local. For this reason let us consider the
variation of ${\rm Im} S_{eff}[\pi]$ in respect to the pion matrix
$U(x)$. We have \cite{DPP2}

\[
\delta\:{\rm Im} S_{eff}[\pi] = -N_c\:\delta\:{\rm Im}\ln\det D
=\frac{iN_c}{2}\Sp\left(\frac{1}{D}\delta D-\frac{1}{D^\dagger}
\delta D^\dagger\right)
\]
\beq
=\frac{iN_c}{2}\Sp\left[(D^\dagger D)^{-1}D^\dagger\delta D
-(DD^\dagger )^{-1}D^\dagger \delta D^\dagger\right].
\la{imsplog1}\eeq
Now one can put in explicit expressions for $D,D^\dagger$ from
\eqs{herm}{ReS}. The aim of this excercise is to get $\dd U$ in the
denominators so that an expansion in this quantity similar to that
of the previous subsection could be used.

Using the Dirac algebra relations (in Euclidean space)

\[
\{\gamma_\mu,\gamma_nu\}=2\delta_{\mu\nu},\;\;\;\;\;
\gamma_\mu^\dagger=\gamma_\mu,\;\;\;\;\;
\gamma_5=\gamma_1\gamma_2\gamma_3\gamma_4=\gamma_5^\dagger,
\]
\beq
\{\gamma_5,\gamma_\mu\}=0,\;\;\;\;\;
\gamma_5^2=\gamma_5,\;\;\;\;\;
\tr(\gamma_5\gamma_\alpha\gamma_\beta\gamma_\gamma\gamma_\delta)
=4\epsilon_{\alpha\beta\gamma\delta},
\la{diracalg}\eeq
one gets after expanding \eq{imsplog1} in powers of $\dd \Ug$
the first non-zero term

\beq
\delta {\rm Im} S_{eff}[\pi] = \frac{iN_c}{48\pi^2}
\int\!d^4x\:\epsilon_{\alpha\beta\gamma\delta}
\Tr\left(\partial_\alpha U^\dagger\partial_\beta U\partial_\gamma
U^\dagger \partial_\delta U\:U^\dagger\delta U\right).
\la{imsplog2}\eeq
It can be easily checked that this expression coincides with the
variation of the Wess--Zumino term written in the form a $d=5$
integral \cite{W}

\[
{\rm Im} S_{eff}[\pi]=\frac{iN_c}{240\pi^2}
\int\!d^5x\:\epsilon_{\alpha\beta\gamma\delta\epsilon}
\Tr\Bigl(U^\dagger\partial_\alpha U\Bigr)
\Bigl(U^\dagger\partial_\beta U\Bigr)
\Bigl(U^\dagger\partial_\gamma U\Bigr)
\Bigl(U^\dagger\partial_\delta U\Bigr)
\Bigl(U^\dagger\partial_\epsilon U\Bigr)
\]
\beq
+\;\;{\rm higher\;\; derivative\;\; terms}.
\la{WZ}\eeq
In fact the integrand in \eq{WZ} is a full derivative, however,
to write it explicitly one would need a parametrization of the
unitary matrix $U$. The expansion of \eq{WZ} starts from the fifth
power of $\pi^A(x)$, and it is non-zero only if $N_f\ge 3$
\cite{WZ}. It is important that, similar to ${\rm Re} S_{eff}$,
the imaginary part is also an infinite series in the derivatives.

The EChL \ur{Seff} or, more generally, the low-energy partition
function \ur{ZNf} from where \eq{Seff} has been derived, is invariant
under vector flavour-singlet transformations. Therefore, there should
be a corresponding conserved N\"other baryon current, $B_\mu$. This
current is associated with the imaginary part of $S_{eff}$ only;
since ${\rm Im} S_{eff}$ is an infinite series in the derivatives
so is the associated Noether current $B_\mu$. For the first
Wess--Zumino term \ur{WZ} the corresponding charge is \cite{W}

\beq
B=-\frac{1}{24\pi^2}\int\!d^3\vec{x}\:\epsilon_{ijk}
\Tr\Bigl(U^\dagger\partial_i U\Bigr)
\Bigl(U^\dagger\partial_j U\Bigr)
\Bigl(U^\dagger\partial_k U\Bigr)
+\;\;{\rm higher\;\; derivative\;\; terms}.
\la{windn}\eeq
The explicitly written term is the winding number of the field
$U(x)$. Let me briefly explain this notion.

If $\pi^A(\vec{x})\rightarrow 0$ at spatial infinity so that
$U(\vec{x})\rightarrow 1$ in all directions, one can say that the
spatial infinity is just one point. \Eq{windn} gives then the winding
number for the mapping of the three-dimensional sphere $S^3$ (to which
the flat $d=3$ space is topologically equivalent when $\infty$ is one
point) to the parameter space of the $SU(N_f)$ group. In case $N_f=2$
the parameter space is also $S^3$ so that the mapping is $S^3\mapsto
S^3$. The topologically non-equivalent mappings $U(\vec{x})$, {\em
i.e.} those which can not be continuously deformed one to another, are
classified by their winding number, an integer analytically given by
\eq{windn}. In case of $N_f> 2$ mathematicians prove that mappings are
also classified by integers given by the same \eq{windn}.

There exists a prejudice that the baryon number carried by quarks in
the external pion field coincides with the winding number of that
field: generally speaking it is not so because of the higher derivative
terms omitted in \eq{windn}. Only if the pion field is spatially large
and slowly varying so that one can neglect the higher derivative
terms in \eq{windn} one can say that the two coincide. Otherwise, for
arbitrary pion fields, the baryon number is not related to the winding
number:  the former may be zero when the latter is unity, and {\em vice
versa}.

To see what is going on here, let us calculate directly the baryon
number carried by quarks in an external time-independent pion field
$U(\vec{x})$ \cite{DPP2}. The definition of the baryon charge operator
in the Minkowski space is

\beq
{\hat B} =\frac{1}{N_c}\int\!d^3\vec{x}\:\bar\psi\gamma_0\psi.
\la{BMink}\eeq
Passing to Euclidean space (which we prefer to work with since
functional integrals are more readily defined in Euclidean) one has
to make a substitution $\bar\psi\rightarrow -i\psi^\dagger,
\gamma_0\rightarrow\gamma_4$, so that

\beq
{\hat B} =-\frac{i}{N_c}\int\!d^3\vec{x}\:\psi^\dagger\gamma_4\psi.
\la{BEucl}\eeq
The baryon charge in the path integral formulation of the theory given
by \eq{Zna} is then

\[
B=\langle \hat B\rangle
=-\frac{i}{N_c}\int\!d^3\vec{x}\langle\psi^\dagger\gamma_4\psi\rangle
=-i\int\!d^3\vec{x}\:\Tr\langle x|\gamma_4(i\dd+iM\Ug)^{-1}|x\rangle
\]
\[
=-i\int\!d^3\vec{x}\:\Tr\langle x_4,\vec{x}|
\frac{1}{i\partial_4+i\gamma_4
\gamma_k\partial_k+iM\gamma_4\Ug}|x_4,\vec{x}\rangle
\]
\[
=-i\int\!d^3\vec{x}\int_{-\infty}^{+\infty}\frac{d\omega}{2\pi}\Tr
\langle \vec{x}|\frac{1}{\omega+iH}|\vec{x}\rangle
\]
\beq
=\Sp\;\theta(-H)\;=\;\;{\rm\bf number\;\; of\;\; levels\;\; with\;\;
E<0}.
\la{B1}\eeq Here

\beq
H=\gamma_4\gamma_k\partial_k+M\gamma_4\Ug
\la{DirHam}\eeq
is the Dirac hamiltoniam in the external time-independent pion field
$U(\vec{x})$ and $\theta$ is a step function.

\Eq{B1} is divergent since it sums up the baryon charge of the whole
negative-energy Dirac continuum. This divergence can be avoided by
subtracting the baryon charge of the free Dirac sea, {\it i.e.} with
the pion field switched out, $H_0=\gamma_4\gamma_k\partial_k
+M\gamma_4$:

\beq
B=-i\int\!d^3\vec{x}\int\frac{d\omega}{2\pi}\Tr
\langle \vec{x}|\frac{1}{\omega+iH}-\frac{1}{\omega+iH_0}|\vec{x}\rangle
=\Sp\left[\theta(-H)-\theta(-H_0)\right].
\la{B2}\eeq
In performing the integration over $\omega$ we have closed the $\omega$
integration contour in the upper semiplane. Had we closed it in the
lower semiplane we would obtain $-\Sp[\theta(H)-\theta(H_0)]$ which is
the same result since
$\Sp[\theta(H)+\theta(-H)-\theta(H_0)-\theta(-H_0)]=0$.

We have thus obtained a most natural result: the baryon charge of
quarks in the external pion field is the number of negative-energy
levels of the hamiltonian \ur{DirHam} (the number of the levels of the
free hamiltonian subtracted).

One can perform the gradient expansion for the baryon number
similarly to that of the real part of the EChL. To that end let
us write

\beq
B=-\int\!d^3\vec{x}\int_{-\infty}^{+\infty}\frac{d\omega}{2\pi}
\Tr\langle x|\frac{H}{\omega^2+H^2}-\frac{H_0}{\omega^2+H_0^2}
|x\rangle
\la{B3}\eeq
where

\beq
H^2=-\partial_k^2+M^2-M\gamma_4(\partial_k\Ug),\;\;\;\;\;
H_0^2=-\partial_k^2+M^2.
\la{H1}\eeq
Calculating the matrix element in the plane-wave basis one gets

\[
B=-\int\!d^3\vec{x}\int\frac{d\omega}{2\pi}\int\frac{d^3\vec{k}}
{(2\pi)^3}\:
\Tr\gamma_4\left[\frac{\vec{\gamma}\cdot(\vec{\partial}+i\vec{k})
+M\Ug}
{\omega^2-(\vec{\partial}+i\vec{k})^2+M^2
-M\vec{\gamma}\cdot(\vec{\partial}\Ug)}\right.
\]
\beq
\left.-\frac{\vec{\gamma}\cdot(\vec{\partial}+i\vec{k})}
{\omega^2-(\vec{\partial}+i\vec{k})^2+M^2}
\right]\cdot 1.
\la{B4}\eeq
For slowly varying fields $U(x)$ \eq{B4} can be expanded in
powers of $\partial\Ug$ and $\partial$ (applied ultimately to
$\Ug$). Because of the $\Tr\gamma_5...$ the first non-zero
contribution arises from expanding the denominator in \eq{B4}
to the third power of $\gamma\cdot (\partial\Ug)$. Integrals
over $\omega$ and $k$ should be explicitly performed. After
some simple algebra one gets

\beq
B=-\frac{1}{24\pi^2}\int\!d^3\vec{x}\:\epsilon_{ijk}\Tr
(\partial_iU^\dagger\partial_jU\partial_kU^\dagger U)\;\;
+\;\;{\rm higher\;\;derivative\;\;terms}
\la{B5}\eeq
coinciding with \eq{windn} derived from the Noether current
corresponding to the Wess--Zumino term \ur{WZ}
\cite{DP7,DHF,MKW,NS}.

It should be stressed that the baryon number carried by quarks in
the background pion field is equal to the topological winding
number of the field only if it is a slowly varying one. The deep
reason for it is the following \cite{DPP2}. Imagine we start from
a pion field $U(x)$ whose winding number is one but whose spatial
size is tending to zero. Such a field would have no impact on the
spectrum of the Dirac hamiltonian \ur{DirHam}: it would remain
the same as that of the free hamiltonian, namely it would have
the upper ($E>M$) and lower ($E<-M$) Dirac continua separated by
the mass gap of $2M$.

We now (adiabatically) increase the spatial size of the pion field
preserving its winding number equal to unity. Since the winding
number is dimensionless this can always be done. At certain
critical spatial size the potential well for quarks formed by
the external pion field is wide enough so that a bound-state
level emerges from the upper continuum. With the increase of
the width of the potential well the bound-state level goes down towards
the lower Dirac continuum. Asymptotically, as one blows up the spatial
size of the pion field (always remaining in the winding number equal
unity sector) the bound-state level travels all the way through the
mass gap separating the two continua and joins the lower Dirac sea --
this is a theorem proven in ref. \cite{DPP2}. At this point one would
discover that there is an extra state close to the lower Dirac
continuum (as compared to the free, that is no-field case).
Therefore, one would say that the baryon number is now unity, --
in correspondence to \eqs{windn}{B5}.

In a general case, however, the baryon number of the quark system
is the number of eigenstates of the Dirac hamiltonian \ur{DirHam}
one bothers to fill in. The role of the winding number of the
background pion field is only to guarantee that, if the spatial
size of the field is large enough, the additional bound-state
 level emerging from the upper continuum is a deep one:
asymptotically it goes all the way to the lower continuum.

\vskip 1true cm
\section{The nucleon}
\setcounter{equation}{0}

\subsection{Physical motivations}

All constituent quark models start from assuming that the nearly
massless light quarks of the QCD lagrangian obtain a non-zero
dynamical quark mass $M\approx 350 - 400\:MeV$. This is due
to the spontaneous breaking of chiral symmetry, its microscopic
driving force being, to my belief, instantons, as explained in
section 2. Even if one does not believe in instantons as the
microscopic mechanism of spontaneous chiral symmetry breaking
one has to admit that once a constituent quark mass is introduced
such quarks {\em inevitably} have to interact with Goldstone
pions. The lagrangian $\bar\psi(i\dd-M)\psi$ is not invariant
under axial rotation
$\psi\rightarrow\exp(i\alpha^A\lambda^A\gamma_5)\psi$,
it is the chiral lagrangian \ur{Zna},

\beq
{\cal L}=\bar\psi(i\dd-M\Ug)\psi,
\la{chirla}\eeq
which is, since the rotation of the quark fields can be
compensated in this lagrangian by renaming of the pion fields.

Expanding $\Ug=\exp(i\pi^A\lambda^A\gamma_5/F_\pi)$ in powers of
the properly normalized pion fields (that is why we have inserted
the $F_\pi$ constant) we see that the dimensionless constant
of the linear coupling of pions to constituent quarks
is, numerically, quite large:

\beq
g_{\pi qq}=\frac{M}{F_\pi}\simeq 4.
\la{coupling}\eeq
I would like to emphasize that this is a model-independent
consequence of saying that quarks get a constituent mass.

Actually, the coupling \ur{coupling} is so strong that one may
wonder how some people manage to get along without taking it into
account.  Not surprisingly, baryon models which do take into
account pion exchange between constituent quarks give much more
realistic predictions than, say, the old simple-minded
Isgur--Karl model (for a review see ref. \cite{G}).

Moreover, at distances between quarks of the order of 0.5 fm
typical for interquark distances inside nucleons, neither the
one-gluon exchange nor the supposed linear potential are as large
as the chiral forces. Therefore, it is worthwhile to investigate
whether the chiral forces alone are able to bind the constituent
quarks inside nucleons. Such approach may or may not be
successful for describing high nucleon excitations where,
according to the standard logic, the confining forces become
crucial. To that we can remark the following:

1. The constituent quark mass is momentum-dependent; the
behaviour of $M(k)$ at low virtualities $k$ may well be divergent
(see section 2) and it may thus play the role of confining forces.
[To my knowledge, this line of thought has not been pursued in the
literature.]

2. Highly excited baryons with large angular momenta, if
understood as chiral solitons, lie on linear Regge trajectories
with a realistic slope related to the $F_\pi$ constant
\cite{DP6,D2}.

Therefore, it may be expected that even highly excited baryons can be
incorporated into the chiral theory. After all, as emphasized by
Witten \cite{W}, the theory of all hadrons can be, in principle,
formulated completely in terms of the EChL. For example, there should
exist an EChL corresponding to the Lovelace--Shapiro dual resonance
amplitudes for $\pi\pi$ scattering exhibiting the correct Regge
behaviour \cite{PV}. A (unsolved) problem is to formulate an
appropriate EChL in the field-theoretic language.

Leaving aside these interesting problems, we concentrate on the lowest
state with baryon number one, {\em i.e.} the nucleon. As mentioned
above the interquark separations in the ground-state nucleon are
moderate (order of 0.5 fm) and it is worthwhile asking whether the
simple EChL \ur{Zna} is capable of explaining the basic properties
of the ground-state nucleon. Notice that the expected typical
momenta of quarks inside the nucleon are of the order of
$M\approx 350-400\:MeV$, that is perfectly inside the domain of
applicability of the low-momentum effective theory \ur{Zna}, according
to its derivation in section 2.

The chiral interactions of constituent quarks in the 3-quark nucleon,
as induced by the effective theory \ur{Zna}, are schematically shown
in Fig.2, where quarks are denoted by solid and pions by dashed lines.
Notice that, since there is no tree-level kinetic energy for pions in
\eq{Zna}, the pion propagation occurs only through quark loops. Quark
loops induce also many-quark interactions indicated in Fig.2 as well.
We see that the emerging picture is, unfortunately, rather far from
a simple one-pion exchange between the constituent quarks: the
non-linear effects in the pion field are not at all suppressed.

\begin{figure}
\centerline{\epsfxsize10.0cm\epsffile{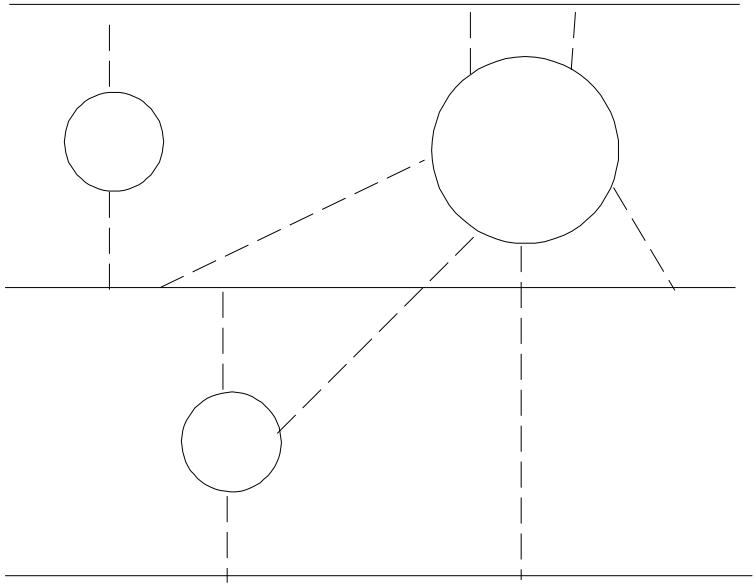}}
\caption[]{Quarks in a nucleon interacting through chiral
fields.}
\end{figure}

At this point one may wonder: isn't the resulting theory as complicated
as the original QCD itself? The answer is no, the effective low-energy
theory is an enormous simplification as compared to the original
quark-gluon theory, because it deals with adequate degrees of freedom.
Let us imagine that one would like to describe `low-energy' properties
of solid states, superconductivity for example. Would working with the
underlying theory (QED) be helpful? Not at all. We know that the
microscopic theory leads, under certain conditions, to the
rearrangement of atoms into a lattice, so that translational symmetry
is spontaneously broken. As a result the Goldstone bosons appear
(here: phonons), and electrons get a dynamical mass different from
the input one. The most important forces are due to phonon exchange
between electrons: in fact they are driving superconductivity in the
BCS theory. Playing with this analogy, nucleon is like a polaron (a
bound state of electrons in the phonon field), rather than a
positronium state in the vacuum. After chiral symmetry is broken
we deal with a `metal' phase rather than with the vacuum one, and one
has to use adequate degrees of freedom to face this new situation.

The instanton vacuum plays the role of the bridge between the
microscopic theory (QCD) and the low-energy theory where one neglects
all degrees of freedom except the Goldstone bosons and fermions with
the dynamically-generated mass. Instantons do the most difficult
part of the job: they explain why atoms in metals are arranged into a
lattice and what is the effective mass of the electron and what is the
strength of the electron-phonon interactions. However, one can
take an agnostic stand and say: I don't care how 350 MeV is obtained
from the microscopic $\Lambda_{{\rm QCD}}$ and why do
atoms form a lattice in the metals: I just know it happens.
To such a person I would advise to take the low-energy theory
\ur{Zna} at face value and proceed to the nucleon.

A considerable technical simplification is achieved in the limit of
large $N_c$. For $N_c$ colours the number of constituent quarks in a
baryon is $N_c$ and all quark loop contributions are also proportional
to $N_c$, see section 3. Therefore, at large $N_c$ one can speak about
a {\em classical self-consistent} pion field inside the nucleon:
quantum fluctuations about the classical self-consistent field will be
suppressed as $1/N_c$. The problem of summing up all diagrams of the
type shown in Fig.2 is thus reduced to finding a classical pion field
pulling $N_c$ massive quarks together to form a bound state.

\subsection{Nucleon mass: a functional of the pion field}

Let us imagine that there is a classical time-independent pion
field which is strong and spatially wide enough to make a  bound-state
level of the Dirac hamiltonian \ur{DirHam} for massive quarks, call its
energy $E_{{\rm level}}$. We fill in this level by $N_c$ quarks in the
antisymmetric state in colour, thus obtaining a baryon number one
state, as compared to the vacuum. The interactions with the background
chiral field are, naturally, colour-blind, so one can put $N_c$ quarks
on the same level; the fact that one has to put them in an
antisymmetric state in colour, {\em i.e.} in a colour-singlet state,
follows from Fermi statistics.

One has to pay for the creation of this trial pion field, however.
Call this energy $E_{{\rm field}}$. Since there are no direct terms
depending on the pion field in the low-momentum theory \ur{Zna}
the only origin of $E_{{\rm field}}$ is the fermion determinant
\ur{Seff} which should be calculated for time-independent field
$U(\vec{x})$. It can be worked out with a slight modification of
section 3. We have \cite{DPP2}:

\[
S_{eff}[\pi]=-N_c\:\ln\;\det\left(\frac{D}{D_0}\right)
\]
\[
=-N_c\;\Sp\left[\ln(i\partial_t+iH)-\ln(i\partial_t+iH_0)\right]
\]
\beq
=-TN_c\int\!\frac{d\omega}{2\pi}\Sp\left[\ln(\omega+iH)
-\ln(\omega+iH_0)\right],
\la{Efi1}\eeq
where $H$ is the Dirac hamiltonian \ur{DirHam} in the stationary pion
field, $H_0$ is the free hamiltonian and $T$ is the (infinite) time
of observation. Using an important relation \footnote{It follows from
taking the matrix trace of the hamiltonian \ur{DirHam}. It should be
kept in mind, though, that such a naive derivation can be potentially
dangerous because of anomalies in infinite sums over levels. However,
it can be checked that in this particular case there are no anomalies
and the naive derivation is correct.}

\beq
\Sp(H-H_0)=0
\la{Spid}\eeq
(telling us that the sum of all energies, with their signs, of
the Dirac hamiltonian \ur{DirHam} is the same as for the free case)
one can integrate in \eq{Efi1} by parts and get

\[
S_{eff}[\pi]\equiv TE_{{\rm field}}
=TN_c\int\!\frac{d\omega}{2\pi}\Sp\left[
\frac{\omega}{\omega+iH}-\frac{\omega}{\omega+iH_0}\right]
\]
\beq
=TN_c\sum_{E_n^{(0)}<0}\left(E_n-E_n^{(0)}\right).
\la{Efi2}\eeq
Going from the first to the second line we have closed the $\omega$
integration contour in the upper semiplane; owing to the trace relation
\ur{Spid} closing it in the lower semiplane would produce the same
result.

We see that the energy cost $E_{{\rm field}}$ one pays for a creation
of the time-independent pion field coincides with the aggregate energy
of the lower Dirac continuum in that field. The energy of the additional
level emerging from the upper continuum, which one has to fill in to
get the baryon number one state, $E_{{\rm level}}$, should be added to
get the total nucleon mass. This simple scheme \cite{DP8,DPP2}
is depicted in Fig.3. Naturally, the mass of the nucleon should be
counted from the vacuum state corresponding to the filled levels of
the free lower Dirac continuum. Therefore, the (divergent) aggregate
energy of the free continuum should be subtracted, as in \eq{Efi2}.

We have thus for the nucleon mass:

\beq{\cal M}_N=\mathrel{\mathop{{\rm min}}\limits_{\{\pi^A(\vec{x})\}}}
\left(N_cE_{{\rm level}}[\pi]+E_{{\rm field}}[\pi]\right).
\la{nuclmass}\eeq

Both quantities, $E_{{\rm level}}$ and $E_{{\rm field}}$, are functionals
of the trial pion field $\pi^A(\vec{x})$. The classical self-consistent
pion field is obtained from minimizing the nucleon mass \ur{nuclmass}
in $\pi^A(\vec{x})$. It is called the {\it soliton} of the
non-linear functional \ur{nuclmass}, hence the {\bf Chiral
Quark-Soliton Model}. An accurate derivation of \eq{nuclmass} from the
path-integral representation for the nucleon-current correlation
function is presented in ref. \cite{DPP2}. It solves the problem
of summing up all diagrams of the type shown in Fig.2 in the
large-$N_c$ limit.

The idea that a sigma model with pions coupled directly
to constituent quarks, can be used to build the nucleon soliton has
been first suggested by Kahana, Ripka and Soni \cite{KRS} and
independently by Birse and Banerjee \cite{BB}. I would call them the
authors of the Chiral Quark-Soliton Model. Technically, however, in
these refs. an additional {\em ad hoc} kinetic energy term for pion
fields has been used, leading to a vacuum instability paradox. The
present formulation of the model has been given in ref. \cite{DP8},
together with the discussion of its domain of applicability and
its physical contents. A detailed theory based on the path
integral approach paving the way for calcultaing nucleon observables
has been presented in ref. \cite{DPP2}.

\begin{figure}
\centerline{\epsfxsize10.0cm\epsffile{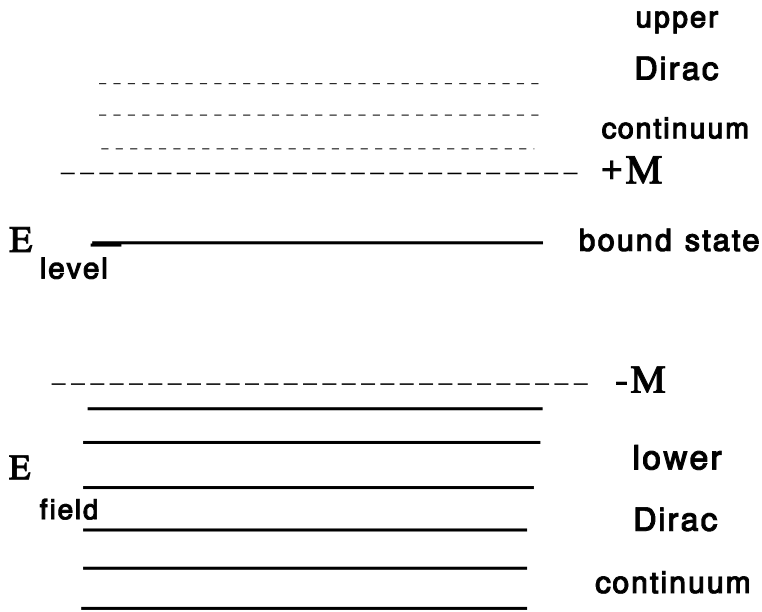}}
\caption[]{Spectrum of the Dirac hamiltonian in trial
pion field. The solid lines present occupied levels.}
\end{figure}

\subsection{Nucleon profile}

To find the classical pion field minimizing the nucleon mass
\ur{nuclmass} one has first of all decide on the symmetry of the
pion field. Had the field been a singlet one would take a spherically-
symmetric ansatz. However, the pion field has flavour indices
$A=1,...,N_f^2-1$. At $N_f=2$ the three components of the pion field
can be married with the three space axes. This is called the
hedgehog ansatz; it is the minimal generalization of spherical symmetry
to incorporate the $\pi^\pm=(\pi^1\pm i\pi^2)/\surd{2}$ and
$\pi^0=\pi^3$ fields:

\beq
\pi^A(\vec{x})=n^AP(r),\;\;\;\;\;n^A=\frac{x^A}{r},\;\;\;\;\;
r=|\vec{x}|,
\la{hedgehog}\eeq
where $P(r)$ is called the soliton profile.

The choice of the ansatz is not at all innocent: baryons corresponding
to different choices would have qualitatively different properties,
see the next subsection. The maximally-symmetric ansatz \ur{hedgehog}
will have definite consequences in applications.

In the $N_f=3$ case there are 8 components of the `pion' field, and
there are several possibilities to marry them to the space axes. The
commonly used ansatz is the `left upper corner' ansatz \cite{W2}:

\beq
U(\vec{x})\equiv \exp\left(i \pi^A(\vec{x})\lambda^A\right)
= \left(\begin{array}{cc}
\exp \left[i(\mbox{\boldmath{$n$}}
\cdot\mbox{\boldmath{$\tau$}}) P(r)\right] &
\begin{array}{c} 0\\ 0\end{array}\\ \begin{array}{cc}0\;\;&\;\;\;0
\end{array}& 1\end{array}\right),\;\;\;\;
\vec{n}=\frac{\vec{x}}{r}.
\la{leftupper}\eeq
As we shall see in the next subsection, the quantization of
rotations for this ansatz leads to the correct spectrum of the lowest
baryons.

Another $SU(3)$ ansatz \cite{Bal} discussed in the literature is
($f,g,h=1,2,3$ are the flavour indices):

\beq
U_{fg}=e^{iP_2/3}\left[\cos P_1\:\delta_{fg}+\left(e^{-iP_2}
-\cos P_1\right)n_fn_g+\sin P_1\:\epsilon_{fgh}n_h\right],
\la{bal}\eeq
where $P_{1,2}(r)$ are spherically-symmetric profile functions.
This ansatz is used to describe strangeness -2 dibaryons \cite{DPPP}.
We shall not consider it here but concentrate on the usual baryons
for which the hedgehog ansatz \ur{hedgehog} or \ur{leftupper} is
appropriate.

Let us first discuss restrictions on the best profile function $P(r)$
which should mimimize the nucleon mass \ur{nuclmass}. What is the
asymptotics of $P(r)$ at large $r$? To answer this question one has
to know the behaviour of $E_{{\rm field}}[\pi]$ for slowly varying pion
fields. Using \eq{Efi2} as a starting point one can work out the
derivative expansion of the functional $E_{{\rm field}}[\pi]$
similar to that for the full EChL, see subsection 3.1. We have
\cite{DP8,DPP2}

\[
E_{{\rm field}}[\pi]=\frac{F_\pi^2}{4}\int\!d^3\vec{x}\:\Tr L_iL_i
-\frac{N_c}{192\pi^2}\int\!d^3\vec{x}\:\left[2\Tr(\partial_iL_i)^2
+\Tr L_iL_jL_iL_j\right]
\]
\beq
+\;\;{\rm higher\;\; derivative\;\; terms},
\;\;\;\;\;L_i=iU^\dagger\partial_iU.
\la{Efi3}\eeq

Substituting here the hedgehog ansatz one gets a functional of the
profile function $P(r)$; varying it one finds the Euler--Lagrange
equation valid for slowly varying profiles, in particular, for
the tail of $P(r)$ at large $r$. It follows from this equation
that $P(r)=A/r^2$ at large $r$. The second contribution to the nucleon
mass, $E_{{\rm level}}[\pi]$, does not alter this derivation since
the bound-state wave function has an exponential, not power behaviour at
large $r$. Actually we get the pion tail inside the nucleon, and the
constant $A$ is related to the nucleon axial constant. This relation
is identical to the one found in the Skyrme model \cite{ANW}:

\beq
g_A=\frac{8\pi}{3}AF_\pi^2.
\la{gA}\eeq
The exponentially decreasing wave function of the bound-state level
does not change this derivation, as well as the Goldberger--Treiman
relation for the pion-nucleon coupling constant,

\beq
g_{\pi NN}=\frac{g_A{\cal M}_N}{F_\pi}
=\frac{8\pi A}{3}\frac{{\cal M}_N}{F_\pi}.
\la{GT}\eeq
Furthermore, it follows from the next four-derivative term in \eq{Efi3}
that the $1/r^4$ correction to $P(r)$ at large $r$ is absent
\cite{DP8,DPP2} !  It means probably that the pion tail inside nucleon
is unperturbed to rather short distances.

The second important question is what should we choose for $P(0)$?
As explained in subsection 3.2, a quantity which guarantees a
deeply-bound state in the background pion field is the winding number
of the field, \eq{windn}. Substituting the hedgehog ansatz into
\eq{windn} one gets

\beq
N_{wind}=-\frac{2}{\pi}\int_0^\infty\!dr\:\sin^2P(r)\frac{dP(r)}{dr}
=-\frac{1}{\pi}\left[P(r)-\frac{\sin 2P(r)}{2}\right]_0^\infty.
\la{windhedge}\eeq
Since $P(\infty)=0$ the way to make this quantity unity is to choose
$P(0)=\pi=3.14...$.

An example of a one-parameter variational function satisfying the above
requirements is \cite{DP8,DPP2}

\beq
P(r)=2\:{\rm arctg}\left(\frac{r_0}{r}\right)^2,\;\;\;\;\;\;
A=2r_0^2.
\la{arctg}\eeq

The Dirac hamiltonian \ur{DirHam} in the hedgehog pion field
\ur{hedgehog} commutes neither with the isospin operator $T$ nor
with the total angular momentum $J=L+S$ but only with their sum
$K=T+J$ called the `grand spin'. The eigenvalue Dirac equations for
given value of $K^2, K_3$ have been derived in refs. \cite{DP8,DPP2}.
Generally speaking, there appears a bound-state level with the
$K^P=0^+$ quantum numbers whose energy can be found from solving the
Dirac equations for two spherically-symmetric functions $j,h$

\be
\frac{dh}{dr}=-Mh\:\sin P+(E_{{\rm level}}+M\:\cos P)\:j,\\
\frac{dj}{dr}+\frac{2}{r}\:j=Mj\:\sin P+(-E_{{\rm level}}
+M\:\cos P)\:h
\la{K0}\ee
with the boundary conditions $h(0)=1,\; j(0)=Cr,\;
h(\infty)=j(\infty)=0$.

These equations determine one of the two contributions to the nucleon
mass, $E_{{\rm level}}$. The second contribution, namely that of
the aggregate energy of the lower Dirac continuum in the trial
pion field, which we have called $E_{{\rm field}}$, can be found
in several different ways. One way is to find the phase shifts in the
lower continuum, arising from solving the Dirac equation for definite
grand spin $K$ \cite{DPP2,DPP3}. Another method is to diagonalize
the Dirac hamiltonian \ur{DirHam} in the so-called Kahana--Ripka
basis \cite{RK} written for a finite-volume spherical box. Both
methods are, numerically, rather involved. There exists a third
(approximate) method \cite{DP8,DPP2} allowing one to make an estimate of
$E_{{\rm field}}$ in a few minutes on a PC. It is based on the
interpolation formula for the EChL, see \eq{2interpol} and
ref. \cite{DPP2} for details.

Let us discuss the qualitative behaviour of $E_{{\rm level}}$ and
$E_{{\rm field}}$ with the soliton scale parameter $r_0$ assuming
for definiteness that the profile is given by \eq{arctg}.

The trial pion field plays the role of the (relativistic) potential
well for massive quarks. The `depth' of this potential well is fixed
by the condition $P(0)=\pi$ and cannot be made infinite: this is
related to the fact that the pion field has the meaning of angles.
The spatial size of the trial pion field $r_0$ plays the role of the
`width' of the potential well. It is well known that in three
dimensions the condition for the appearance of a bound state is
$MVr_0^2>{\rm const}$ where $V$ is the depth of well and `const' is a
numerical constant of the order of unity depending on the concrete
shape of the potential well. In our case $V\approx M$, so the condition
that the bound state appears is $Mr_0\sim 1$. Therefore, at small
sizes $r_0$ there is no bound state for the Dirac hamiltonian
\ur{DirHam}, so that $E_{{\rm level}}$ coincides with the border of the
upper Dirac continuum, $E_{{\rm level}}=+M$. At certain critical value
of $r_0$ a weakly bound state emerges from the upper continuum. [For
the concrete ansatz \ur{arctg} the threshold value is $r_0M\simeq
0.5$.] As one increases $r_0$ the bound state goes deeper and $E_{{\rm
level}}$ monotonously decreases. At very large spatial sizes,
$r_0\rightarrow\infty$, $E_{{\rm level}}$ approaches the lower
continuum, its difference from $-M$ falling as $1/r_0^2$ \cite{DPP2}.
The behaviour of $E_{{\rm level}}$ as function of $r_0$ is plotted in
Fig.4.

The monotonous decrease of $E_{{\rm level}}$ with the increase of $r_0$
is a prerogative of the trial pion field with winding number 1.
Had it been zero, $E_{{\rm level}}$ would first go down and then
start to go up, asymptotically joining back the {\em upper} continuum.
In the case of $N_{wind}=-1$ the bound state would travel in the
opposite direction: from the lower towards the upper continuum. At
$N_{wind}=n$ as much as $n$ levels would emerge, one by one, from the
upper continuum and travel all the way through the mass gap towards the
lower one. For the trial pion field of the hedgehog form all these
things happen exclusively for states with grand spin
$K=0$ \cite{DPP2}.

Turning now to $E_{{\rm field}}$ we first notice that for large
spatial sizes of the trial pion field one can use the first term
in the derivative expansion for $E_{{\rm field}}$, see \eq{Efi3}.
On dimension grounds one immediatelly concludes that
$E_{{\rm field}}\sim F_\pi^2r_0$ for large $r_0$, {\em i.e.} is
infinitely linearly rising in $r_0$. At small $r_0$ a slightly more
complicated analysis \cite{DPP2} shows that  $E_{{\rm field}}\sim
r_0^3$. On the whole, $E_{{\rm field}}$ is a monotonously rising
function of $r_0$ shown in Fig.4.

\begin{figure}
\centerline{\epsfxsize10.0cm\epsffile{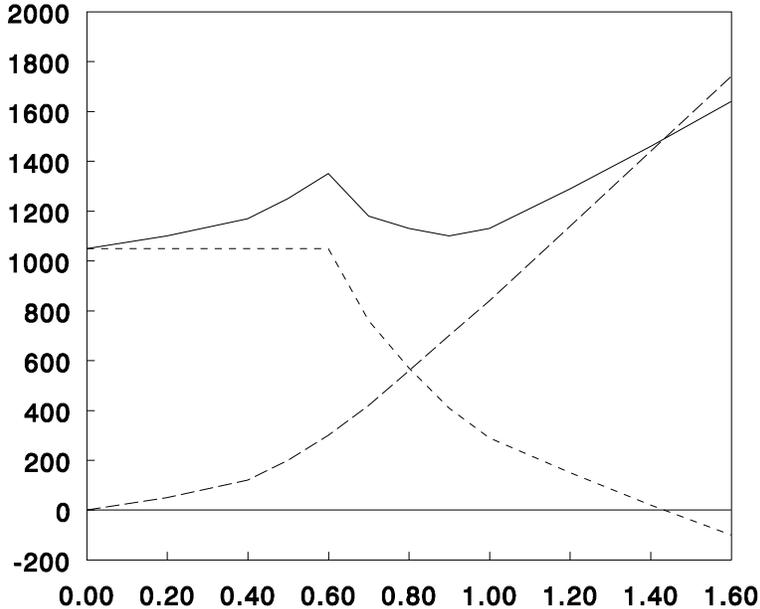}}
\caption[]{The classical nucleon mass and its constituents
as function of the soliton size. The short-dash line shows
$3E_{{\rm level}}$, the long-dash line shows $E_{{\rm field}}$, the
solid line is their sum, ${\cal M}_N$.}
\end{figure}

The nucleon mass, ${\cal M}_N=3E_{{\rm level}}+E_{{\rm field}}$
(for $N_c=3$) is also plotted in Fig.4 taken from
refs. \cite{DP8,DPP2}. One observes a non-trivial minimum for
${\cal M}_N$ corresponding to $r_0\simeq 0.98/M\simeq 0.57\:fm$.
This is, phenomenologically, a very reasonable value, since from
\eqs{gA}{GT} one immediatelly gets
$g_A\simeq 1.15\;{\rm versus}\;1.25\;{\rm (exp.)}$ and
$g_{\pi NN}\simeq 13.6\;{\rm versus}\;13.5\;{\rm (exp.)}$.
The nucleon mass appears to be ${\cal M}_N\simeq 1100\:MeV$
with $E_{{\rm level}}\simeq 123\:MeV$,
$E_{{\rm field}}\simeq 730\: MeV$.
Note that the `valence' quarks (sitting on the bound-state level)
come out to be very strongly bound:  their wave function falls
off as $\exp(-r/0.6\:fm)$, and about $2/3$ of the quark mass
$M\simeq 350\:MeV$ is eaten up by interactions with the classical
pion field. Relativistic effects are thus essential.

Though the nucleon bound state appears to be somewhat higher
than the free-quark threshold, $3M\simeq 1050\:MeV$,
there are several known corrections to it which all seem to be
negative.  The largest correction to the nucleon mass is due to
taking into account explicitly the one-gluon exchange between both
`valence' and `sea' quarks; this correction is $O(N_c)$ as is the
nucleon mass itself. Numerically, it turns out to be about
$-200\:MeV$ \cite{DJP} and seems to move the nucleon mass just into
the right place \footnote{There exist also numerous {\em quantum}
corrections to the nucleon mass of different origin, which are of
the order of $O(N_c^0)$.  Unfortunately, it is difficult today to
treat them in a systematic fashion; see, however, the next
subsection.}.

We see thus that the `valence' quarks in the nucleon get bound
by a self-consistent pion field whose energy is given just by
the aggregate energy of the negative Dirac continuum distorted
by the presence of the external field. This picture of the nucleon
interpolates between the old non-relativistic quark models (which
would correspond to a shallow bound-state level and an undistorted
negative continuum) and the Skyrme model (which would correspond to a
spatially very large pion soliton so that the bound-state level would
get close to the lower continuum and the field energy $E_{{\rm field}}$
would be given just by a couple of terms in its derivative expansion).
The reality is somewhere in between: the bound-state level is a deep
one but not as deep as to say that all the physics is in the lower
Dirac continuum.

Idelogically, this picture of the nucleon at large $N_c$ is
somewhat similar to the Thomas--Fermi picture of the atom at large
$Z$. In that case quantum fluctuations of the self-consistent
electrostatic field binding the electrons are suppressed by large
$Z$, however corrections go as powers of $Z^{-2/3}$. Therefore, the
Chiral Quark-Soliton Model is in a slightly better position in
respect to quantum corections than the Thomas--Fermi approximation.

Apart from using the large $N_c$ approximation (which is in fact
just a technical device needed to justify the use of the classical
pion field) the Chiral Quark-Soliton Model makes use of the
small algebraic parameter $(M\bar\rho)^2$ where $\bar\rho$ is the
average size of instantons in the vacuum. This $\bar\rho$ is, roughly,
the size of the constituent quark, while the size of the nucleon is,
parametrically, $1/M$. The fact that the constituent quark picture
works so well in the whole hadron physics finds its explanation in
this small numerical parameter being due to the relative diluteness
of the instanton vacuum, which in its turn is related to the
`accidentally' large number (11/3) in the asymptotic freedom law
\cite{DP1,DPW}. The small parameter $(M\bar\rho)^2\ll 1$ makes it
possible to use only quarks with dynamically generated mass and
chiral fields as the only essential degrees of freedom in the
range of momenta $k\sim M\ll 1/\bar\rho$, and that is exactly
the range of interest in the nucleon binding problem.

The above numerics have been obtained from the interpolation
formula for $E_{{\rm field}}$ \cite{DP8,DPP2}. Exact calculations
of $E_{{\rm field}}$ performed in \cite{DPP3} as well as taking
more involved profiles with three variational parameteres did not
lead to any significant changes in the numerics.

Following refs. \cite{DP8,DPP2} there had been many
calculations of the nucleon mass and of the `best' profile using
various regularization schemes and parameteres of the chiral model,
see \cite{Review} for a review. The effective theory derived from
the instanton vacuum comes with an intrinsic ultraviolet cutoff, in
the form of a momentum dependence of the constituent quark mass,
$M(k)$. It can be shown on general grounds that this is a rapidly
falling function at momenta of the order of the inverse average
instanton size, $1/\bar\rho \approx 600\, {\rm MeV}$. However, the
present `state of the art' does not allow one to determine this
function accurately at all values of momenta -- to do so, one
would need a very detailed understanding of the instanton vacuum.
This places certain restrictions on the kinds of quantities which
can sensibly be computed using the effective theory.  Those are
either finite ones, which do not require an UV cutoff at all, or
quantities at most logarithmically divergent. Both type of
quantities are dominated by momenta much smaller than the UV
cutoff, $k \ll 1/\bar\rho$, so one can compute them mimicking the
fall--off of $M(k)$ by an external UV cutoff
$\Lambda\simeq 1/\bar\rho$, using  some regularization scheme.
Fortunately, almost all nucleon observables belong to these two
classes. The uncertainty related to the details of the ultra-violet
regularization leads to a 15-20\% numerical uncertainty of the
results, and that is the expected accuracy of the model today.

\subsection{Quantum numbers of baryons}

The picture of the nucleon outlined in the previous subsection is
``classical'': the quantum fluctuations of the self-consistent
pion field binding $N_c$ quarks are totally ignored.
Among all possible quantum corrections to the nucleon mass
a special role belongs to the zero modes. Fluctuations of the pion
field in the direction of the zero modes cannot be considered small,
and one has to treat them exactly. Zero modes are always related to
continuous symmetries of the problem at hand. In our case there are
3 zero translational modes and a certain number of zero rotational
modes.  The latter determine the quantum numbers of baryons;
it is here that the hedgehog (or whatever) symmetry of the ansatz
taken for the self-consistent pion field becomes crucial.

A general statement is that if the chiral field $U_{{\rm cl}}(\vec{x})$
minimizes the nucleon mass functional \ur{nuclmass}, a field
corresponding to rotated spatial axes, $x_i\rightarrow O_{ij}
x_j$, or to a unitary-rotated matrix in flavour space,
$U_{{\rm cl}}\rightarrow RU_{{\rm cl}}R^\dagger$, has obviously
the same classical mass. This is because the functional \ur{nuclmass}
to be minimized is isotropic both in flavour and ordinary spaces.

Specifically for the hedgehog ansatz [see \eq{hedgehog} for the flavour
$SU(2)$ and \eq{leftupper} for the $SU(3)$] any spatial rotation is
equivalent to a flavour rotation. We show it for a more complicated
case of $SU(3)$.  Indeed, the space-rotating $3\times 3$ matrix
$O_{ij}$ can be written as

\beq
O_{ij}=\frac{1}{2}\Tr (S\tau_i S^\dagger\tau_j)
\la{Oij}\eeq
where $S$ is an $SU(2)\;$ $\;2\times 2$ matrix and $\tau_i$ are the
three Pauli matrices. One can immediatelly check that $O_{ij}$ are real
orthogonal 3-parameter matrices with
$O_{ij}O_{kj}=\delta_{ik}$ and $O_{ij}O_{ik}=\delta_{jk}$, as it
should be.

When one rotates the space putting $n_i^\prime=O_{ij}n_j$ the
$2\times 2$ matrix standing in the left upper corner of the ansatz
\ur{leftupper} can be written as

\[
\exp \left[i(\mbox{\boldmath{$n^\prime$}}
\cdot\mbox{\boldmath{$\tau$}}) P(r)\right]
=\cos P(r)+ i(\mbox{\boldmath{$n^\prime$}}
\cdot\mbox{\boldmath{$\tau$}})\sin P(r)
\]
\beq
=S\left[\cos P(r)+ i(\mbox{\boldmath{$n$}}
\cdot\mbox{\boldmath{$\tau$}})\sin P(r)\right]S^\dagger.
\la{cossin}\eeq
Therefore, if one consideres the hedgehog ansatz \ur{leftupper}
rotated {\em both} in flavour and usual spaces, the latter can be
completely absorbed into the former one:

\beq
R\:U_{{\rm cl}}(O\vec{x})\:R^\dagger
=\tilde R\:U_{{\rm cl}}(\vec{x})\:\tilde R^\dagger
\la{equivrot}\eeq
with

\beq
\tilde R= R\left(\begin{array}{cc}S&\begin{array}{c}0\\0\end{array}\\
\begin{array}{cc}0 &\;\;0 \end{array}& 1\end{array}\right).
\la{absorb}\eeq

For that reason it is sufficient to consider rotations only in the
flavour space. Hence there are 3 zero rotational modes in the $SU(2)$
and 8-1=7 in the $SU(3)$ flavour case. The rotation of the form
$R=\exp(i\alpha\lambda^8)$ commutes with the left-upper-corner
ansatz and therefore does not correspond to any zero mode. This will
have important consequences in getting the correct spectrum of
hyperons.

The general strategy \cite{DPP2} is to consider a slowly rotating
ansatz

\beq
\tilde U(\vec{x},t)=R(t)\:U_{{\rm cl}}(\vec{x})\:R^\dagger (t)
\la{rotating}\eeq
and to expand the energy of the bound-state level and of the negative
Dirac continuum in `right' ($\Omega_A$) and `left'
($\tilde\Omega_A$) angular velocities

\beq
\Omega_A=-i\Tr(R^\dagger \dot R\lambda^A),\;\;\;\;\;
\tilde\Omega_A=-i\Tr(\dot R R^\dagger\lambda^A),\;\;\;\;\;
\Omega^2=\tilde\Omega^2=2\Tr\dot R^\dagger\dot R.
\la{angvel}\eeq

Taking into account only the lowest terms in the time derivatives
of the rotation matrix $R(t)$ one gets \cite{DPP2,B} the following
form of the rotation lagrangian:

\beq
L^{rot} = \frac{1}{2}I_{AB}\Omega_A\Omega_B
-\frac{N_c}{2\sqrt{3}}\Omega_8.
\la{Lrot}
\eeq
Here $I_{AB}$ is the $SU(3)$ tensor of the moments of inertia,
\beq
I_{AB}=\frac{N_c}{4}\int\frac{d\omega}{2\pi}\:\Tr\left(
\frac{1}{\omega+iH}\lambda^A \frac{1}{\omega+iH}\lambda^B\right)
\la{IAB}\eeq
where the $\omega$ integration contour should be drawn {\em above}
the bound-state energy $E_{{\rm level}}$ to incorporate the `valence'
quarks.

The appearance of a linear term in $\Omega_8 $ is an important
consequence of the presence of an extra bound-state level emerging from
the upper Dirac continuum, which fixes the baryon charge to be unity.
We remind the reader that in the Skyrme model this linear term arises
from the Wess-Zumino term \cite{Gua}. For simplicity we have written
unregularized moments of inertia, though \eq{IAB} should be regularized
in some way, see {\em e.g.} \cite{B}.

Owing to the left-upper-corner ansatz for the static soliton
(being essentially $SU(2)$) the tensor $I_{AB}$ is diagonal and depends
on two moments of inertia, $I_{1,2}$:

\beq
I_{AB}=\left\{ \begin{array}{cc} I_1\:\delta_{AB},\;&\;A,B=1,2,3,\\
            I_2\:\delta_{AB},\;&\;A,B=4,5,6,7,\\ 0,\;&\;A,B=8.
\end{array}\right.
\eeq
Therefore, the rotational lagrangian \ur{Lrot} can be rewritten as

\beq
L^{rot}=\frac{I_1}{2} \sum_{A=1}^{3}\Omega_A^2  +
\frac{I_2}{2} \sum_{A=4}^{7}\Omega_A^2 -\frac{N_c}{2\sqrt{3}}\Omega_8.
\la{rotlagr}
\eeq

To quantize this rotational Lagrangian  one can use the canonical
quantization procedure, same as in the Skyrme model
\cite{Gua,DP7,MNP,Chemtob,JW}.  Introducing eight angular momenta
canonically conjugate to `right' angular velocities $\Omega_A$,

\beq J_A=\frac{\partial L^{rot}}{\partial \Omega_A},
\la{moments}\eeq
and writing the hamiltonian as
\beq
H^{rot}=\Omega_A J_A-L^{rot}
\la{rothamgen}\eeq
one gets
\beq
H^{rot}=\frac{1}{2I_1}\sum_{A=1}^3
J_A^2+\frac{1}{2I_2}\sum_{A=4}^7 J_A^2
\la{Ham1}\eeq
with the additional quantization prescription following from \eq{moments},

\beq
J_8=-\frac{N_c}{2\surd{3}}=-\frac{\surd{3}}{2}.
\la{quantiz}\eeq

In the Skyrme model this quantization rule follows from the Wess-Zumino
term. In our approach it arises from filling in the
bound-state level, i.e. from the `valence' quarks. It is known to
lead to the selection rule: not all possible spin and $SU(3)$
multiplets are allowed as rotational excitations of the $SU(2)$
hedgehog. \Eq{quantiz} means that only those $SU(3)$ multiplets are
allowed which contain particles with hypercharge $Y=1$; if the number
of particles with $Y=1$ is denoted as $2J+1$, the spin of the allowed
$SU(3)$ multiplet is equal to $J$.

Therefore, the lowest allowed $SU(3)$ multiplets are:

\begin{itemize}
\item octet with spin 1/2 (since there are {\em two} baryons in the octet
with $Y=1$, the $N$)
\item decuplet with spin 3/2 (since there are {\em four} baryons in the
decuplet with $Y=1$, the $\Delta$)
\item anti-decuplet with spin 1/2 (since there are {\em two} baryons in the
anti-decuplet with $Y=1$, the $N^*$)
\end{itemize}
The next are 27-plets with spin $1/2$ and $3/2$ but we do not
consider them here.

We see that the lowest two rotational excitations are exactly the lowest
baryon multiplets existing in reality. The third predicted multiplet,
the anti-decuplet, contains exotic baryons which cannot be made of
three quarks, most notably an exotic $Z^+$ baryon having spin 1/2,
isospin 0 and strangeness +1. A detailed study of the anti-decuplet
performed recently in \cite{DPPol} predicts that such a baryon
can have a mass as low as 1530 MeV and be very narrow. Several
experimental searches of this exotic baryon are now under way.

It is easy to derive the splittings between the centers of the
multiplets listed above. For the representation $(p,q)$ of the $SU(3)$
group one has

\beq
\sum\limits_{A=1}^8 J_A^2 = \frac13[p^2+q^2+pq + 3(p+q)],
\eeq
therefore the eigenvalues of the rotational hamiltonian
\ur{Ham1}) are

\beq
E_{(p,q)}^{rot} = \frac1{6I_2} [p^2+q^2+pq + 3(p+q)]
+\left(\frac1{2I_1} - \frac1{2I_2} \right) J(J+1)
- \frac{3}{8I_2}.
\la{rotenerg}\eeq
We have the following three lowest rotational excitations:

\beqr
(p,q)=(1,1), & J=1/2:&\; \mbox{octet, spin 1/2}, \\
(p,q)=(3,0), & J=3/2:&\; \mbox{decuplet, spin 3/2}, \\
(p,q)=(0,3), & J=1/2:&\; \mbox{anti-decuplet, spin 1/2} \, .
\eeqr
The splittings between the centers of these multiplets are determined
by the moments of inertia, $I_{1,2}$:

\beq
\Delta_{10-8} =
E_{(3,0)}^{rot} - E_{(1,1)}^{rot} =\frac{3}{2I_1},
\la{octdecspl}\eeq
\beq
\Delta_{{\overline{10}} - 8} =
E_{(0,3)}^{rot} - E_{(1,1)}^{rot} =\frac{3}{2I_2}.
\la{oct-antidecspl}\eeq

The appropriate rotational wave functions describing
members of these multiplets are given by Wigner finite-rotation
functions $D^{8,10,{\overline{10}}}(R)$ \cite{B,DPPol}.

When dealing with the flavour $SU(3)$ case neglecting the strange
quark mass $m_s$ is an oversimplification. In fact, it is easy to
incorporate $m_s\neq 0$ in the first order. As a result one gets
very reasonable splittings inside the $SU(3)$ multiplets, as well as
mass corrections to different observables \cite{B,Review,DPPol}.

In general, the idea that all light baryons are rotational excitations
of one object, the `classical' nucleon, leads to numerous relations
between properties of the members of octet and decuplet, which follow
purely from symmetry considerations and which are all satisfied up
to a few percent in nature. The $SU(3)$ symmetry by itself says
nothing about the relation between different multiplets, of course.
Probably the most spectacular is the Guadagnini formula
\cite{Gua} which relates splittings inside the decuplet with those in
the octet,

\beq
8(m_{\Xi^*}+m_N)+3m_\Sigma=11m_\Lambda+8m_{\Sigma^*},
\la{Guad}\eeq
which is satisfied with better than one-percent accuracy!

\vskip .5true cm
\underline{${\bf N_f=2}$ {\bf case}}
\vskip .5true cm

If one is interested in baryons predominantly `made of' $u,d$ quarks,
the flavour group is $SU(2)$ and the quantization of rotations is
more simple.

In this case the rotational lagrangian is just

\beq
L^{rot}=\frac{I_1}{2} \sum_{i=1}^{3}\Omega_i^2
=\frac{I_1}{2} \sum_{A=1}^{3}\tilde\Omega_A^2,
\la{rotlagr2}
\eeq
where the `right' ($\Omega_i$) and `left' ($\tilde\Omega_A$) angular
velocities are defined by \eq{angvel}. This is the lagrangian for the
spherical top: the two sets of angular velocities have the meaning
of those in the `lab frame' and `body fixed frame'. The quantization
of the spherical top is well known from quantum mechanics. One has
to introduce two sets of angular momenta, $S_i$ (canonically conjugate
to $\Omega_i$) and $T_A$ (conjugate to $\tilde \Omega_A$). Both sets
of operators act on the coordinates of the spherical top, say,
the Euler angles. It will be more convenient for us to say that
the coordinates of the spherical top are just the entries of the
unitary matrix $R$ defining its finite-angle rotation \cite{DPP2}.

The angular momenta operators $S_i,\; T_A$ act on $R$ as
generators of right (left) multiplication,

\beq
e^{i(\alpha S)}Re^{-i(\alpha S)} = R\:e^{i(\alpha\sigma)},
\la{S}\eeq
\beq
e^{i(\alpha S)}Re^{-i(\alpha S)} = e^{-i(\alpha\tau)}\:R,
\la{T}\eeq
and satisfy the commutation relations

\[
[T_A,T_B]=i\epsilon_{ABC}T_C ,\;\;\;\;\;[S_i,S_j]=i\epsilon_{ijk}S_k,
\]
\beq
[T_A,S_i]=0,\;\;\;\;\;\;(T_A)^2=(S_i)^2.
\la{comm}\eeq
A realization of these operators is

\[
S_i=R_{pq}\left(\frac{\sigma_i}{2}\right)_{kq}
\frac{\partial}{\partial R_{pq}},
\]
\beq
T_A=-\left(\frac{\tau_A}{2}\right)_{pk}R_{kq}
\frac{\partial}{\partial R_{pq}}.
\la{realiz}\eeq

The rotational hamiltonian is

\beq
H^{rot}=\Omega_i S_i-L^{rot}=\tilde\Omega_A T_A-L^{rot}
=\frac{S_i^2}{2I_1} =\frac{T_A^2}{2I_1}.
\la{rotham2}\eeq

Comparing the definition of the generators \urs{S}{T} with the
ansatz \ur{rotating} we see that $T_A$ is the flavour (here: isospin)
operator and $S_i$ is the spin operator, since the former acts to the
left from $R$ and the latter acts to the right.

The normalized eigenfunctions of the mutually commuting operators
$S_3,\;T_3$ and $S^2=T^2$ with eigenvalues $S_3,\;T_3$ and
$S(S+1)=T(T+1)$ are \cite{DPP2}

\beq
\Psi^{(S=T)}_{T_3S_3}(R)=\sqrt{2S+1}(-1)^{T+T_3}
D^{(S=T)}_{-T_3S_3}(R)
\la{eigenfu}\eeq
where $D(R)$ are Wigner finite-rotation matrices. For example, in the
$S=T=1/2$ representation $D^{1/2}_{pq}(R)=R_{pq}$, {\em i.e.} coincides
with the unitary matrix $R$ itself.

The rotational energy is thus

\beq
E^{rot}=\frac{S(S+1)}{2I_1} =\frac{T(T+1)}{2I_1}
\la{roten2}\eeq
and is $(2S+1)^2=(2T+1)^2$-fold degenerate. The wave functions
\ur{eigenfu} describe at $S=T=1/2$ four nucleon states (proton,
neutron, spin up, spin down) and at $S=T=3/2$ the sixteen
$\Delta$-resonance states, the splitting between them being

\beq
m_{\Delta}-m_N=\frac{3}{2I_1} = O(N_c^{-1})
\la{split2}\eeq
(coinciding in fact with the splitting between the centers of decuplet
and octet in the more general $SU(3)$ case, see \eq{octdecspl}).

It is remarkable that the nucleon and its lowest escitation, the
$\Delta$, fits into this spin-equal-isospin scheme, following from
the quantization of the hedgehog rotation. Moreover,
since $N$ and $\Delta$ are, in this approach, just different
rotational states of the same object, the `classical nucleon',
there are certain relations between their properties. These relations
are identical to those found first in the Skyrme model \cite{ANW}
since they follow from symmetry considerations only and do not depend
on concrete dynamics which is of course different in the naive
Skyrme model. For example, one gets for the dynamics-independent ratio
of magnetic moments and pion couplings \cite{ANW}

\[
\frac{\mu_{\Delta N}}{\mu_p-\mu_n}=\frac{1}{\surd{2}}\simeq 0.71
\;\;\;\;vs.\;\;\;\;0.70\pm 0.01\;\;({\rm exp.}),
\]
\beq
\frac{g_{\pi N\Delta}}{g_{\pi NN}}=\frac{3}{2}=1.5
\;\;\;\;vs.\;\;\;\;1.5\pm 0.12\;\;({\rm exp.}).
\la{rels2}\eeq

I should mention that there might be interesting implications of
the `baryons as rotating solitons' idea to nuclear physics. The
low-energy interactions between nucleons can be viewed as interactions
between spherical tops depending on their relative orientation
$R_1R_2^\dagger$ in the spin-isospin spaces \cite{D3,DM}. It leads
to an elegant description of $NN$ and $N\Delta$ interactions in a
unified fashion, and it would be very interesting to check its
experimantal consequences (as far as I know this has not been done
yet). A nuclear medium is then a medium of interacting quantum
spherical tops with extremely anizotropic interactions depending
on relative orientations of the tops both in the spin-isospin and
in ordinary spaces.

This unconvential point of view is strongly supported by the
observation \cite{D3} that one can get the correct
value of the so-called symmetry energy of the nucleus,
$25\:MeV\cdot(N-Z)^2/A$, with the coefficient $25\:MeV$ appearing as
$1/8I_1$ where $I_1$ is the $SU(2)$ moment of inertia; from the
$\Delta - N$ splitting \ur{split2} one finds
$I_1\simeq (200\;MeV)^{-1}$. I do not know whether the language of
spherical tops is fruitful to describe ordinary nuclear matter
(probably it is but nobody tried), however it is certainly useful
to address new questions, for example whether nuclear matter at high
densities can be in a strongly correlated antiferromagnet-type
phase \cite{DM}.

Finally, let us ask what the next rotational excitations could be?
If one restricts oneself to only two flavours, the next state should
be a (5/2, 5/2) resonance; in the three-flavour case the third
rotational excitation is the anti-decuplet with spin $1/2$, see above.
Why do not we have any clear signal of the exotic (5/2, 5/2) resonance?
The reason is that the angular momentum $J=5/2$ is numerically
comparable to $N_c=3$. Rotations with $J\approx N_c$ cannot be
considered as slow: the centrifugal forces deform considerably the
spherically-symmetric profile of the soliton field \cite{DP6,BR};
simultaneously at $J\approx N_c$ the radiation of pions by the rotating
body makes the total width of the state comparable to its mass
\cite{DP6,D2,Dorey}. In order to survive strong pion radiation the
rotating chiral solitons with $J\ge N_c$ have to stretch into
cigar-like objects; such states lie on linear Regge trajectories
with the slope $\alpha^\prime\approx 1/8\pi^2F_\pi^2$ \cite{DP6,D2}.

The situation, however, might be somewhat different in the
{\it three}-flavour case. First, the rotation is, roughly
speaking, distributed among more axes in flavour space, hence
individual angular velocities are not neccessarily as large as when we
consider the two-flavour case with $J=5/2$. Actually, the $SU(2)$
baryons with $J=5/2$ belong to a very high multiplet from the $SU(3)$
point of view. Second, the radiation by the soliton includes now $K$
and $\eta$ mesons which are substantially heavier than pions, and
hence such radiation is suppressed. Actually, the anti-decuplet
seems to have moderate widths \cite{DPPol} and it is worthwhile
searching for the predicted exotic states.

\subsection{Some applications}

There exists by now a rather vast literature studying baryon observables
in the Chiral Quark-Soliton Model \footnote{Sometimes the model has
been called `the solitonic sector of the Nambu--Jona-Lasinio model'.
I have many objections to this title. First, Vaks and Larkin
have suggested independently and at the same time a 4-fermion
model to illustrate how symmetry can be dynamically broken in field
theory. Therefore, in any case I would call it the VL/NJL model.
Second, both VL \cite{VL} and NJL \cite{NJL} were talking about
nucleons as fundamental fermions (there were no quarks in 1961),
and this is rather far from what we consider now. Third, as
discussed in section 2, the instanton-induced interactions in
contrast to the {\em ad hoc} 4-fermion interactions correctly reproduce
the symmetries: $U_A(1)$ is explicitly broken, while in the $N_c=2$
case they possess a more wide $SU(4)\times U(4)$ symmetry \cite{DP5};
at $N_f>2$ they necessarily are $2N_f$-fermion interactions and not
at all 4-fermion. Fourth, instanton-induced interactions provide
a natural UV cutoff, as given by the momentum-dependence of the
constituent quark mass. For the success of the Chiral Quark-Soliton
Model it is extremely important that this UV cutoff is much larger
than the constituent quark mass itself (actually squared),
meaning that the size of the constituent quark is much less than
the size of the nucleon. In an {\em ad hoc} 4-fermion model
having no obvious relation to QCD one has to impose the UV cutoff
by hands. Fifth, in an arbitrarily introduced 4-quark interaction
model there are no {\em a priori} reasons to freeze out all degrees
of fredom except the chiral ones. Meanwhile, if one includes
the $\sigma$ field into the minimization of the nucleon mass
the soliton collapses \cite{S,WT}. In short, when one knows results
coming from instantons, it is possible to mimic {\em some} of them
by imposing certain rules of the game with the 4-quark
interactions. But why then should it be called the `NJL model'?}.
Baryon formfactors (electric, magnetic and axial), mass splittins,
the nucleon sigma term, magnetic moments, weak decay constants,
tensor charges and many other characteristics of nucleons and
hyperons have been calculated in the model. I address the reader to
an extensive review \cite{Review} on these matters.

Here I would like to point out several developments of the Chiral
Quark-Soliton Model interesting from the theoretical point of view.
The list below is, of course, very subjective.

The study of the spin content of the nucleon in the model has been
pioneered by Wakamatsu and Yoshiki \cite{WY}. They showed that the
fraction of the nucleon spin carried by the spin of quarks is
about 50\% (and could be made less): the rest is carried
by the interquark orbital moment, the Dirac sea contribution to it
being quite essential.

An important question is $1/N_c$ corrections to baryon observables.
These can be classified in two groups: one comes, for example, from
meson loops and is therefore accompanied by an additional small
factor $\sim 1/8\pi^2$, the second arises from a more accurate
account for the quantization of the zero rotational modes. The
second-type corrections are not accompanied by small loop factors,
and may be quite substantial: after all in the real world $N_c=3$
so a 30\% correction is not so small. Such corrections for
certain quantities have been fished out in refs. \cite{WW,CGPPWW}
in the two-flavour case and in ref. \cite{BPG} for three flavours.
These corrections work in a welcome direction: they lower the
fraction of nucleon spin carried by quark spins and increase the
flavour non-singlet axial constants.

A recent development of the model deals with the parton
distributions in nucleon. This topic deserves however a special
subsection.

\subsection{Nucleon structure functions}

The distribution of quarks, antiquarks and gluons, as measured
in deep inelastic scattering of leptons, provides us probably with the
largest portion of quantitative information about strong interactions.
Until recently only the {\em evolution} of the structure functions
from a high value of the momentum transfer $Q^2$ to even higher values
has been succesfully compared with the data. This is the field of
perturbative QCD, and its success has been, historically, essential in
establishing the validity of the QCD itself. However, the initial
conditions for this evolution, namely the leading-twist distributions
at a relatively low normalizaion point, belong to the field of
non-perturbative QCD. If we want to understand the vast amount of data
on unpolarized and polarized structure functions we have to go into
non-perturbative physics.

The Chiral Quark-Soliton Model presents a non-perturbative approach
to the nucleons, and it is worthwhile looking into the parton
distributions it predicts. Contrary to several models of nucleons
on the market today, it is a relativistic field-theoretical model.
This circumstance is of crucial importance when one deals with parton
distributions. It is only with a relativistic field-theoretical model
one can preserve general properties of parton distributions such as

\begin{itemize}
\item  relativistic invariance,
\item  positivity of parton distributions,
\item  partonic sum rules which hold in full QCD.
\end{itemize}

There are two seemingly different ways to define parton distributions.
The first, which I would call the Fritsch--Gell-Mann definition, is
a nucleon matrix element of quark bilinears with a light-cone
separation between the quark $\psi$ and $\bar\psi$ operators.
According to the second, which I would call the Feynman--Bjorken
definition, parton distributions are given by the number of partons
carrying a fraction $x$ (the Bjorken variable) of the nucleon
momentum in the nucleon infinite-momentum frame. See Feynman's book
\cite{F} for the discussion of both definitions. In perturbative QCD
only the Fritsch--Gell-Mann definition has been exploited as one has no
idea how to write down the nucleon wave function in the
infinite-momentum frame, which is necessary for the Feynman--Bjorken
definition.

Despite the apparent difference in wording, it has been shown for the
first time, whithin the field-theoretical Chiral Quark-Soliton Model,
that the two definitions are, in fact, equivalent and lead to
identical working formulae for computing parton distributions:
in ref. \cite{SF1} the first definition has been adopted while
in ref. \cite{SF2} the second was used. The deep reason for that
equivalence is that the main hypothesis of the Feynman--Bjorken
parton model, namely that partons transverse momenta do not grow with
$Q^2$ \cite{F}, is satisfied in the model.

Let me point out some key findings of refs. \cite{SF1,SF2}. \\

(i) {\bf Classification of quark distributions in $N_c$} \\

Since the nucleon mass is $O(N_c)$ all parton distributions are
actually functions of $xN_c$. Combining this fact with the known
large-$N_c$ behaviour of the integrals of the distributions over
$x$ one infers that all distributions can be divided in `large'
and `small'. The `large' distributions are, for example, the
unpolarized singlet and polarized isovector distributions, which are of
the form

\beq
D^{{\rm large}}(x)\sim N_c^2f(xN_c),
\la{large}\eeq
where $f(y)$ is a stable function in the large-$N_c$ limit. On the
contrary, the polarized singlet and unpolarized isovector distributions
give an example of `small' distributions, having the form

\beq
D^{{\rm small}}(x)\sim N_cf(xN_c).
\la{small}\eeq
One, indeed, observes in experiment that `large' distributions are
substantially larger than the `small' ones. \\

(ii) {\bf Antiquark distributions}\\

In the academic limit of a very weak mean pion field in the nucleon
the Dirac continuum reduces to the free one (and should be subtracted
to zero) while the bound-state level joins the upper Dirac continuum.
In such a limit there are no antiquarks, while the distribution of
quarks becomes $q(x)=N_c^2\delta(xN_c-1)$. In reality there is a
non-trivial mean pion field which {\em a)} creates a bound-state level,
{\em b)} distorts the negative-energy Dirac continuum. As the result,
the above $\delta$-functiuon is smeared significantly, and a non-zero
antiquark distribution appears.

An inevitable consequence of the relativistic invariance is that the
bound-state level makes a {\em negative}-definite contribution to the
antiquark distribution \footnote{This is also true for any nucleon
model with valence quarks, for example for any variant of bag models.
Bag models are essentially non-relativistic, so they fail to resolve
this paradox. In order to cure it, one has to take into account
contributions to parton distributions from {\em all} degrees of freedom
involved in binding the quarks in the nucleon. That can be consistently
done only in a relativistic field-theoretical model, like the one
under consideration.}. The antiquark distribution becomes positive
only when one includes the contribution of the Dirac continuum.
Numerically, the antiquark distribution appears to be sizeable
even at a low normalization point, in accordance with phenomenology.\\

(iii) {\bf Sum rules}\\

The general sum rules holding in full QCD are automatically satisfied
in the Chiral Quark-Soliton Model: in refs. \cite{SF1,SF2} the validity
of the baryon number, isospin, total momentum and Bjorken sum rules
has been checked. In fact, it is for the first time that nucleon
parton distributions at a low normalization point have been
consistently calculated in a relativistic model preserving all general
properties.  \\

(iv) {\bf Smallness of the gluon distribution}\\

As many times stressed in these lectures, the whole approach of the
Chiral Quark-Soliton Model is based on the smallness of the algebraic
parameter $(M\bar\rho)^2$ where $\bar\rho$ is the average size of
instantons in the vacuum. This $\bar\rho$ is the size of  the
constituent quark, while the size of the nucleon is, parametrically,
$1/M$. Computing parton distribution in the model one is restricted
to momenta $k \ll 1/\bar\rho\approx 600\:MeV$, so that the
internal structure of the constituent quarks remains unresolved.
There are no gluons in the nucleon at this resolution scale;
indeed, the momentum sum rule is satisfied with quarks and
antiquarks only.

However, when one moves to the resolution scale of 600 MeV or
higher, the constituent quarks cease to be point-like, and that is
at this scale that a non-zero gluon distribution emerges. Having
a microscopic theory of how quarks get their dynamical masses
one can compute the non-perturbative gluon distribution in the
constituent quarks \footnote{Steps in that direction has been
taken in refs. \cite{DPW,BPW}.}. What can be said on general
grounds is that the fraction of momentum carried by gluons is of
the order of $(M\bar\rho)^2\approx 1/3$, which seems to be the
correct portion of gluons at a low normalization point of about 600
MeV where the normal perturbative evolution sets in.\\

(v) {\bf Comparison with phenomenology}\\

There are several parametrizations of the nucleon parton
distributions at a relatively low normalization point, which,
after their perturbative evolution to higher momentum transfer
$Q^2$, fit well the numerous data on deep inelastic scattering.
The most daring (and convenient for our purpose) parametrization is
that of Gl\"uck, Reya {\em et al.} \cite{GR1,GR2} who pushed the
normalization point for their distributions to as low as 600 MeV,
starting from the perturbative side. In refs. \cite{SF1,SF2} parton
distributions following from the Chiral Quark-Soliton Model have
been compared with those of refs. \cite{GR1,GR2}. There seems to be
a good qualitative agreement though the constituent quark and
antiquark distributions appear to be systematically `harder' than
those of \cite{GR1,GR2}. This deviation is to be expected since
the structure of the constituent quarks themselves has not been
yet taken into account, see above.

\section{Conclusions}
\setcounter{equation}{0}

The Chiral Quark-Soliton Model is a simple and elegant
reduction of the full-scale QCD at low energies, however
preserving its main ingredients, namely spontaneous chiral symmetry
breaking, and the appearance of the dynamical (or constituent)
quark mass. Personally, I prefer the word `dynamical': first,
because it is, indeed, dynamiclly generated, second, because it is
momentum-dependent.

The momentum dependence of the dynamical quark
mass $M(k)$ is the key to understanding why the notion of
constituent quarks have worked so remarkably well over 30 years
in hadron physics. The point is, the scale $\Lambda$ at which
the function $M(k)$ falls off appears to be much larger than
$M(0)$; the former parameter determines the size of constituent
quarks while the latter parameter determines the size of hadrons.
These two distinctive scales come neatly from instantons, as
described in section 2.

The Chiral Quark-Soliton Model fully exploits the existence of the
two distinctive scales: it is because of them it makes sense to
restrict oneself to just two degrees of freedom in the nucleon problem,
namely, to massless or nearly massless (pseudo) Goldstone pions and to
the constituent quarks with a momentum-dependent dynamical mass
$M(k)$. The scale $\Lambda$ actually plays the role of the physical
ultra-violet cutoff for the low-energy theory; its domain of
applicability is thus limited to the range of momenta
$k\sim M < \Lambda$. This is precisely the domain of interest for
the nucleon binding problem.

A technical tool simplifying considerably the nucleon problem is
the use of the large $N_c$ logic. At large $N_c$ the nucleon is
heavy, and one can speak of the classical self-consistent pion
field binding the $N_c$ valence quarks of the nucleon together.
The classical pion field (the soliton) is found from minimizing the
energy of the bound-state level plus the aggregate energy of the
lower Dirac continuum in a trial pion field. The valence
quarks (sitting on the bound-state level) appear to be strongly
bound by the classical pion field.

By quantizing the slow rotations of the soliton field in flavour
and ordinary spaces one gets baryon states which are rotational
excitations of the static `classical nucleon'. The classification
of the rotational excitations depends on the symmetry properties of
the soliton field, but not on the details of dynamics. Taking
the hedgehog ansatz one gets the following lowest baryon
multiplets: octet with spin 1/2, decuplet with spin 3/2 (these are,
indeed, the lowest multiplets observed in nature) and antidecuplet
with spin 1/2. This last multiplet contain baryons with exotic
quantum numbers (in the sense that they cannot be composed of only
three quarks); some of them are predicted to be relatively light
and narrow resonances, and it would be of great interest to search
for such states.

By sayng that all lightest baryons are nothing but rotational
excitations of the same object, the `classical nucleon', we get
many relations between members of baryon multiplets which are all
realized with astonishing accuracy in nature. Especially successful
are predictions which do not depend on dynamical quantities (like
the values of moments of inertia) but follow from symmetry
considerations only, and are therefore shared, {\em e.g.}, by the
Skyrme model. Predictions of the Chiral Quark-Soliton Model which
do depend on concrete dynamics are, in general, also in good
accordance with reality: the typical accuracy for numerous baryon
observables computed in the model is about 15-20\%, coinciding
with the expected theoretical accuracy of the model. To get a
better accuracy one needs a better understanding of the underlying
QCD vacuum and of the resulting effective low-energy theory.
The developed theory of the instanton vacuum of QCD seems to do the
job of explaining the hadron world pretty well already, however if
one wants to improve the accuracy of predictions one has to make
the theory more precise.

To my knowledge, the Chiral Quark-Soliton Model is the only
relativistic field-theoretical model of the nucleon on the market
today, and this advantage of the model becomes crucial when one
turns to the numerous parton distribution functions.
It is impossible to get a consistent description of parton
distributions satisfying positivity and sum rules restrictions,
without having a relativistic theory at hand and without taking
into account the complete set of forces which bind quarks together.
The leading-twist parton distributions computed so far in the
Chiral Quark-Soliton Model refer to a very low normalization point
where the structure of the constituent quarks is not resolved yet.
Nevertheless, they seem to be in qualitative agreement with
parametrizations of the DIS data at low $Q^2$ though, not
unnaturally, they appear to be more `hard'.

I think that it is the field of parton distributions where the
Chiral Quark-Soliton Model will be used most of all in the near
future. We know how to (perturbatively) evolve parton
distributions fron high to still higher values of $Q^2$ but we do
not really understand how to explain the initial conditions for
that evolution, that is the leading-twist parton distributions at a
low normalization point.  This is where a relativistic model
satisfying all general requirements could be of great use.  Also in
the years to come there will be much experimental activity
involving numerous spin and off-forward parton distributions, as
well as non-leading-twist distributions. Practically nothing is
known about these numerous distributions from the theoretical side,
and the predictions of the Chiral Quark-Soliton Model can be very
valuable, see refs. \cite{BPW,PP,PPPBGW} for the first
predictions.

\vskip 1true cm
{\bf Acknowledgements}
\vskip .5true cm

I wish to gratefully acknowledge the remarkable efforts by the
Peniscola School organizers, Manuel Asorey and Antonio Dobado.

I am most grateful to my colleagues Victor Petrov, Pavel Pobylitsa,
Maxim Polyakov for numerous discussions over the past years of the
topics presented here and to Klaus Goeke and Christian Weiss for a
fruitful collaboration.

\vskip 1true cm

\end{document}